\documentclass[aps, prb, 10pt, twocolumn, superscriptaddress, notitlepage, floatfix, longbibliography]{revtex4-2}
\usepackage{amsmath, amssymb, amsthm, mathtools}
\usepackage{dsfont}
\def\id{\mathds{1}}

\def\Z{\mathds{Z}}
\def\N{\mathds{N}}
\def\bO{\mathcal{O}}
\usepackage{braket}
\usepackage{xcolor}
\usepackage{graphicx}
\usepackage{hyperref}
\usepackage{indentfirst}

\usepackage{tikz}
\usetikzlibrary{decorations.markings}
\tikzset{arrowmid/.style={decoration={markings, mark= at position 0.2 with {\arrow{stealth}}}, postaction={decorate}}}

\newcommand{\hkh}[1]{\left\{ #1 \right\}}
\newcommand{\fkh}[1]{\left[ #1 \right]}
\newcommand{\kh}[1]{\left( #1 \right)}
\newcommand{\euler}{\,\mathrm{e}}
\newcommand{\ii}{\mathrm{i}}
\newcommand*\diff{\mathop{}\!\mathrm{d}}

\DeclareMathOperator{\Tr}{Tr}
\DeclareMathOperator{\sgn}{sgn}
\DeclarePairedDelimiter\abs{\lvert}{\rvert}%
\DeclarePairedDelimiter\norm{\lVert}{\rVert}%
\makeatletter
\let\oldabs\abs
\def\abs{\@ifstar{\oldabs}{\oldabs*}}
\let\oldnorm\norm
\def\norm{\@ifstar{\oldnorm}{\oldnorm*}}
\makeatother
\newcommand{\overbar}[1]{\mkern 1.5mu\overline{\mkern-1.5mu#1\mkern-1.5mu}\mkern 1.5mu}

\def\U{\mathrm{U}(1)}
\def\H{\mathcal{H}}

\begin{document}
\title{Universal contributions to charge fluctuations in spin chains at finite temperature}
\author{Kang-Le Cai}
\author{Meng Cheng}
\affiliation{Department of Physics, Yale University, New Haven, Connecticut 06511, USA}

\begin{abstract}
At finite temperature, conserved charges undergo thermal fluctuations in a quantum many-body system in the grand canonical ensemble. The full structure of the fluctuations of the total $\U$ charge $Q$ can be succinctly captured by the generating function $G(\theta)=\langle\euler^{\ii \theta Q}\rangle$. For a 1D translation-invariant spin chain in the thermodynamic limit, the magnitude $\abs{G(\theta)}$ scales with the system size $L$ as $\ln \abs{G(\theta)}=-\alpha(\theta)L+\gamma(\theta)$, where $\gamma(\theta)$ is the scale-invariant contribution and may encode universal information about the underlying system. In this work, we investigate the behavior and physical meaning of $\gamma(\theta)$ when the system is periodic. We find that $\gamma(\theta)$ only takes nonzero values at isolated points of $\theta$, which is $\theta=\pi$ for all our examples. In two exemplary lattice systems, we show that $\gamma(\pi)$ takes quantized values when the $\U$ symmetry exhibits a specific type of 't Hooft anomaly with other symmetries. In other cases, we investigate how $\gamma(\theta)$ depends on microscopic conditions (such as the filling factor) in field theory and exactly solvable lattice models.
\end{abstract}

\maketitle

\section{Introduction}
At finite temperature, the state of a quantum many-body system with $\U$ symmetry is conveniently described by the thermal density matrix $\rho=\frac{1}{Z}\euler^{-\beta (H-\mu Q)}$ in the grand canonical ensemble, where $\beta$ is the inverse temperature and $\mu$ is the chemical potential. It is well-known that even though the total charge $Q$ is not conserved in the ensemble, it has a well-defined average with fluctuations suppressed in the thermodynamic limit. Higher moments of $Q$ are also suppressed as the distribution approaches a Gaussian one. Our main objective in this work is to show that the charge fluctuations may contain universal information about the system, especially when the $\U$ charge-conservation symmetry exhibits certain types of 't Hooft anomaly (usually a mixed anomaly with other global symmetries). To extract this information, it is most useful to consider the following generating function (also known as the full counting statistics~\cite{FCS1, FCS2, FCS3, FCS4})
\begin{equation}
    G(\theta)=\braket{\euler^{\ii\theta Q}} =\Tr \euler^{\ii\theta Q}\rho.
\end{equation}
In (1+1)d, for a system of size $L$, it is expected that $G(\theta)$ takes the following form
\begin{align}
    \ln \abs{G(\theta)} = -\alpha(\theta)L + \gamma(\theta)+\cdots,
\end{align}
for large $L$, and the phase factor is defined as
\begin{equation}
    \omega(\theta)=\frac{G(\theta)}{\abs{G(\theta)}}.
\end{equation}
In general, the value of $\alpha(\theta)$ is sensitive to microscopic details and is thus non-universal. On the other hand, $\gamma(\theta)$ and $\omega(\theta)$ are expected to encode universal information about many-body systems. 

Another motivation to study $G(\theta)$ comes from its connection with the disorder parameter in the (2+1)d ground state of a gapped Hamiltonian~\cite{WangPRB2021, WuSP2021, Chen:2022upe, CuomoPRL2022}. Suppose $A$ is a subregion of the (2+1)d system, and denote by $Q_A$ the total charge inside $A$. We define the disorder parameter as $\braket{\euler^{\ii\theta Q_A}}$, where the expectation value now is taken with respect to the ground state. When the ground state is gapped and preserves the $\U$ symmetry, one expects for a region $A$ of linear size $L_A$
\begin{equation}
  \ln\abs{\braket{\euler^{\ii\theta Q_A}}} =-\alpha_1(\theta)L_A+\gamma_1(\theta)+\bO(1/L_A),
\end{equation}
where $\gamma_1(\theta)$ is known as the topological disorder parameter. To make connections to the (1+1)d discussion more explicit, we note that in many cases the reduced density matrix can be well-approximated by a thermal state of a (quasi-)local Hamiltonian acting on degrees of freedom localized at the boundary of $A$, which is effectively a (1+1)d system. If the (2+1)d bulk is a nontrivial symmetry-protected topological (SPT) state, the symmetries should act anomalously in the effective boundary theory. Thus the computation of $\braket{\euler^{\ii \theta Q_A}}$ in the (2+1)d system is reduced to $G(\theta)$ in an effective (1+1)d system, and the topological disorder parameter $\gamma_1(\theta)$ is identified with $\gamma(\theta)$.

In this work, we study the behavior and the physical interpretations of $\gamma(\theta)$ and $\omega(\theta)$.  A key question to address is to what extent the values of $\gamma(\theta)$ are truly universal, i.e. unaffected by small changes to the Hamiltonian. Additionally, we explore the physical significance of these universal, quantized values.

First, we present a general computation of $G(\theta)$ when the (1+1)d system can be described by a conformal field theory (CFT), possibly with a topological defect.  In the absence of such a defect, we find that $\gamma(\theta)$ effectively counts the degeneracy of the defect operator of the corresponding $\U$ symmetry transformation. It only takes nonzero values at isolated values of $\theta$, as some of the degeneracies are enforced by the mixed anomaly between $\U$ and other global symmetries. 

We then compute $\gamma(\theta)$ and $\omega(\theta)$ in the presence of a $\U$ topological defect for a $c=1$ free boson CFT. This computation is essential for comparing the CFT results with those obtained from a lattice model. It turns out that both $\gamma(\theta)$ and $\omega(\theta)$ are sensitive to the $\U$ defect, which can be understood as a consequence of the 't Hooft anomaly of the $\U\times\U$ symmetry of the free boson CFT. 

Next, we examine $\gamma(\theta)$ and $\omega(\theta)$ in lattice models to understand how they behave at high temperature, beyond the scope of the CFT description. We also address the question as to whether the value of $\gamma(\theta)$ is quantized or not. Our main finding is that in the presence of certain types of mixed anomaly between $\U$ and other global symmetry, $\gamma(\pi)$ takes quantized values for a general thermal state of any symmetric local Hamiltonian. Hence $\gamma(\theta)$ provides a way to characterize the 't Hooft anomaly in 1D quantum systems through a nonlocal observable at finite temperature.

We establish this result for two types of systems: 
\begin{enumerate}
    \item A translation-invariant spin-1/2 chain with on-site $\mathrm{O}(2)$ symmetry, where $\mathrm{O}(2)$ and lattice translation have Lieb-Schultz-Mattis (LSM) anomaly. We show that $\gamma(\pi)=\ln 2$ (for even system size).
    \item A translation-invariant spin-1/2 chain with anomalous $\mathrm{O}(2)\times\Z_2$ symmetry, where the $\U$ symmetry is not on-site. We show that 
    \begin{align}
        \gamma(\pi)
        =\begin{cases}
            \ln 2 & N\equiv 0\pmod{4} \\
            \ln \sqrt{2} & N\equiv \pm 1 \pmod{4}
        \end{cases}.
    \end{align} 
\end{enumerate}
While the 't Hooft anomalies in these two systems appear quite different, they share a common feature: a $\pi$ flux of the $\U$ symmetry carries a projective representation protected by other symmetries. This type of 't Hooft anomaly is often referred to as the ``type-III" anomaly in literature~\cite{deWildPropitius:1995cf}. The type-III structure of the anomaly turns out to be crucial for the quantization of $\gamma(\pi)$.

We compute $G(\theta)$ in two spin chain lattice models at finite temperature: \textbf{1}. the spin-1/2 XX chain, which has the $\mathrm{O}(2)$ LSM anomaly, and \textbf{2.} the Levin-Gu spin chain with the $\mathrm{O}(2)\times\Z_2$ mixed anomaly. Both models can be solved exactly using Jordan-Wigner transformations, and the results for $\gamma(\pi)$ indeed agree with the expectations. We match the full results for $G(\theta)$ with the CFT predictions at low energy, which requires the understanding of the continuum limit as the CFT with insertions of emanant symmetry defects. We also show explicitly in these examples that if the symmetry is reduced so the anomaly is no longer of type-III, $\gamma(\pi)$ is not quantized anymore and can vary continuously depending on microscopic parameters, such as the filling factor.

More generally, we provide arguments for the quantization of $\gamma(\pi)$ based on the matrix product representation of the density matrix for both cases mentioned above, generalizing an argument in \cite{Zang}. As a by-product, we find that in both cases the $\U$ symmetry can be reduced to the $\Z_2$ subgroup without affecting the results, as the type-III anomaly structure is preserved.

Finally, we discuss possible crossover behavior of $\gamma(\theta)$ in finite-size systems. We use the antiferromagnetic Ising model to illustrate how $\gamma(\theta)$ approaches the infinite-size value as we vary system size and temperature.

\section{Charge fluctuations in a CFT}

\begin{figure}[ht]
\centering
\begin{tikzpicture}[scale=0.7]
    \draw[red, thick] (-1.5,-2) arc (180:360:1.5 and 0.5);
    \node[red] at (0,-2) {$\euler^{\ii\theta Q}$};
    \draw[red, dashed] (-1.5,-2) arc (180:0:1.5 and 0.5);
    \draw[thick] (0,0) ellipse (1.5 and 0.5);
    \draw[arrowmid, thick] (-1.5,-4) -- (-1.5,0);
    \node at (-1.8,-3.5) {$\tau$};
    \draw[thick] (1.5,0) -- (1.5,-4);
    \draw[arrowmid, thick] (-1.5,-4) arc (180:360:1.5 and 0.5);
    \node at (-1,-4.7) {$x$};
    \draw[dashed] (-1.5,-4) arc (180:0:1.5 and 0.5);

    \draw[thick, ->] (2.5,-2) to[bend right=20] (3.5,-2);

    \draw[red, thick] (6,-0.5) -- (6,-4.5);
    \node[red] at (6.5,-2) {$\euler^{\ii\theta Q}$};
    \draw[thick] (6,0) ellipse (1.5 and 0.5);
    \draw[arrowmid, thick] (4.5,-4) -- (4.5,0);
    \node at (4.2,-3.5) {$\tau$};
    \draw[thick] (7.5,0) -- (7.5,-4);
    \draw[arrowmid, thick] (4.5,-4) arc (180:360:1.5 and 0.5);
    \node at (5,-4.7) {$x$};
    \draw[dashed] (4.5,-4) arc (180:0:1.5 and 0.5);
\end{tikzpicture}
\caption{Path integral representation of $G(\theta)$. Left cylinder: In Eq.\eqref{cft_eq_G1}, $\euler^{\ii\theta Q}$ represents a defect at fixed imaginary time. Right cylinder: After a spacetime rotation, $\euler^{\ii\theta Q}$ becomes a spatial defect, as described by Eq.\eqref{cft_eq_G2}.}
\label{fig_cft_spacetimerotation}
\end{figure}

We begin by considering the case where the (1+1)d system is described by a CFT, at least at the energy scale of interest to us. More precisely, we assume that at low energy, the Hamiltonian $H$ can be approximated by 
\begin{equation}
    H\approx \frac{2\pi v}{L}H_{\rm CFT}+E_0(L),
\end{equation}
where $L$ is the length of the system, $H_{\rm CFT}=L_0+\overbar{L}_0$ is the dimensionless CFT Hamiltonian on a unit circle $S^1$, and $v$ is the velocity. $E_0(L)$ is the  ground state  energy. We define a rescaled temperature $\tilde{\beta}=\frac{2\pi v\beta}{L}$.

In this case, we will show below that $\gamma(\theta)$ can be related to vacuum degeneracy in the defect sector. To see this, recall that we need to evaluate
\begin{equation}
    G(\tilde{\beta},\theta)=\braket{\euler^{\ii\theta Q}}=\frac{\Tr \euler^{\ii\theta Q}\euler^{-\tilde{\beta} H_{\rm CFT}}}{\Tr \euler^{-\tilde{\beta} H_{\rm CFT}}}=\frac{Z(\tilde{\beta},\theta)}{Z(\tilde{\beta},0)}.\label{cft_eq_G1}
\end{equation}
Here we have defined the (twisted) partition function
\begin{equation}
    Z(\tilde{\beta},\theta)=\Tr \euler^{\ii\theta Q}\euler^{-\tilde{\beta} H_{\rm CFT}}.
\end{equation}
We can think of $Z(\tilde{\beta},\theta)$ as the Euclidean path integral of the theory on a spacetime torus, with the operator $\euler^{i\theta Q}$ inserted at a fixed time, as illustrated by the left panel of Fig.~\ref{fig_cft_spacetimerotation}. Since by definition the CFT is invariant under spacetime rotations, one can perform a $\pi/2$ rotation on the torus, swapping space and time, and the partition function should remain the same (this is a special case of the modular invariance). Now after rotation we have the partition function of the CFT at temperature $1/\tilde{\beta}$, twisted by a $\euler^{\ii\theta Q}$ defect as shown in the right panel of Fig.~\ref{fig_cft_spacetimerotation}. Therefore
\begin{equation}
\begin{split}
Z(\tilde{\beta},0)&=Z(1/\tilde{\beta},0),\\
    Z(\tilde{\beta},\theta)&=Z_\theta(1/\tilde{\beta}).
\end{split}    
\end{equation}
Here we define 
\begin{equation}
    Z_\theta(\tilde{\beta})=\Tr_{\theta} \euler^{-\tilde{\beta} H_{\rm CFT}},
\end{equation}
where $\Tr_\theta$ means that we perform the trace in the Hilbert space of the CFT with a $\euler^{\ii\theta Q}$ defect. As a result, we find
\begin{equation}
    G(\theta)= \frac{Z_\theta(1/\tilde{\beta})}{Z_0(1/\tilde{\beta})}.\label{cft_eq_G2}
\end{equation}

In the limit of large $L$, and consequently small $\tilde{\beta}$, the modular transformed theory is at low temperature, and the partition function can be approximated by keeping only the ground state contribution. Therefore we obtain
\begin{equation}
    G(\theta)\approx d_\theta \euler^{-\frac{h_\theta}{\tilde{\beta}}}=d_\theta \euler^{-\frac{h_\theta}{2\pi v \beta}L}.
    \label{G-cft-1}
\end{equation}
Here $d_\theta$ is the ground state degeneracy in the defect sector. $h_\theta$ is the (dimensionless) energy of the highest-weight state in the defect sector, which is identified as the scaling dimension of the $\U$ defect operator. 

For a rational CFT, the presence of a $\U$ symmetry implies that the CFT has $\U_{k}\times \U_{-{k}}$ Kac-Moody algebra, with $k$ being the level, and for $\theta\in[0,\pi]$, $h_\theta$ is given by the charged Cardy formula~\cite{Dyer:2017rul, Hosseini:2020vgl, FanPRL2023}:
\begin{equation}
    h_\theta=k\theta^2.
\end{equation}
Therefore,
\begin{equation}
    \gamma(\theta)=\ln d_{\theta}.
\end{equation}
We now recall that, in general, an anomaly in the $\U$ symmetry implies that a certain quantum number of the topological defect must change as $\theta$ continuously varies from $0$ to $2\pi$. This is essentially equivalent to Laughlin's flux-insertion argument. As a consequence, the ground state must become degenerate at some value of $\theta$ where levels cross. Thus the typical behavior of $\gamma(\theta)$ is to remain at 0, except at a few isolated points where it jumps to a finite value.

We note that the argument can be easily generalized to other symmetries. In fact, Eq.~\eqref{G-cft-1} holds without any modification, except that $h_\theta$ should be replaced by the conformal weight of the corresponding defect operator. Essentially equivalent result was found in the context of topological disorder parameter in a (2+1)d gapped state in Ref.~\cite{Chen:2022upe}.

A useful generalization is to consider CFT equipped with a topological defect. For this purpose, instead of a general CFT, we specialize to the example of a $c=1$ Luttinger liquid (or free compact boson CFT). It is arguably the simplest CFT with $\U$ symmetry, which will be most relevant for our lattice model examples below. In general, such a CFT has two $\U$ symmetries, generated by charges $Q_m$ and $Q_w$, respectively. The Hamiltonian has two parts: $H_{\rm CFT}=H_0+H_{\rm osc}$, where $H_{\rm osc}$ is the energy of the oscillator modes, which do not contribute to the global charge fluctuations. The ``zero mode" energy $H_0$ takes the following form:
\begin{equation}
    H_0=\frac{1}{2} \left(\frac{Q_m^2}{R^2} + R^2Q_w^2\right).
\end{equation}
Here $R$ is the radius of the compact boson. In addition to the $\U_m\times\U_w$ symmetry, at a generic radius the CFT also has a charge-conjugation symmetry which acts as $Q_m\rightarrow -Q_m, Q_w\rightarrow -Q_w$, so together the full symmetry is $[\U_m\times\U_w]\rtimes \Z_2$. 

Consider the CFT with a defect corresponding to $\euler^{2\pi\ii \eta_m Q_m}\euler^{2\pi\ii\eta_w Q_w}$ with $\eta_m,\eta_w\in[0,1)$. The effect of the defect is to change the quantization conditions of the charges. The energy levels (the zero mode contributions) become 
\begin{equation}
    H_0(\eta_m, \eta_w)=\frac{1}{2R^2}\kh{Q_m+\eta_w}^2 + \frac{R^2}{2}\kh{Q_w+\eta_m}^2.
\end{equation}
For our purpose, it suffices to consider the case $\eta_m=0$, and to compute $\braket{\euler^{\ii\theta Q_m}}$. Details of the derivation can be found in Appendix~\ref{cft}. Evaluating the partition functions, we find that
\begin{equation}
  \alpha(\theta)= \frac{R^2}{4\pi v\beta}[\theta]_\pi^2,
\end{equation}
where we define
  \begin{equation}
  [\theta]_\pi=\begin{cases}
        {\theta} & 0\leq \theta<\pi \\
        \theta-2\pi & \pi\leq \theta\leq 2\pi
    \end{cases}.
\end{equation}
As expected on general grounds, $\alpha(\theta)$ is ``local" and insensitive to global boundary condition. Notice that $\alpha(\theta)$ exhibits a cusp at $\theta=\pi$, which appears to be a general feature of CFTs with a $\U$ symmetry. A similar cusp was found in the coefficient of the leading term in the $\U$ disorder parameter in a (1+1)d Luttinger liquid or a (2+1)d Fermi liquid~\cite{Tan:2019axb, Jiang:2022tmb, CaiFL}.

The full expression of $G(\theta)$ is found to be
\begin{equation}
    G(\theta)=
    \begin{cases}
        \euler^{-\ii \eta_w[\theta]_\pi }\euler^{-\alpha(\theta)L} & \theta\neq \pi\\
        2\cos(\eta_w\pi)\euler^{-\alpha(\pi)L} & \theta=\pi
    \end{cases}.
    \label{cft-G-eta}
\end{equation}
The ``universal" contribution $\gamma(\theta)$ is given by
\begin{equation}
    \gamma(\theta)=
    \begin{cases}
        0 & \theta\neq \pi\\
        \ln \abs{2\cos \eta_w\pi} & \theta=\pi
    \end{cases}.
    \label{gamma-with-defect}
\end{equation}
Notice that in general $\abs{2\cos\eta_w\pi}$ is not an integer. For $\eta_w=1/2$, one finds that $\gamma(\pi)$ diverges, and $G(\pi)=0$. 

Unlike $\alpha(\theta)$, the prefactors in $G(\theta)$ [i.e., $\gamma(\theta)$ and the phase factor $\omega(\theta)$] only depend on the defect parameter $\eta_w$, and have no dependence on other quantities that are sensitive to microscopic details, such as $v$ and $R^2$. In other words, $\gamma(\theta)$ and $\omega(\theta)$ appear to be robust against small changes to the theory. 

However, an important caveat in this argument for the robustness of $\gamma(\theta)$ and $\omega(\theta)$ is the dependence of $\eta_w$ on the microscopic physics. It has been understood that certain microscopic conditions, such as the filling factor, enter the low-energy theory as background topological defects required by anomaly matching~\cite{ChengPRX2016, MetlitskiPRB2018, ChoPRB2017, JianPRB2018}. As an example, consider a $c=1$ Luttinger liquid in a lattice system with $\U$ filling factor $\nu$ (i.e., the average charge per unit cell is $\nu$). The low-energy physics of a system with periodic boundary condition of $N$ unit cells should be described by the Luttinger liquid with a defect $\eta_w=-N\nu$~\cite{Cheng:2022sgb, MetlitskiPRB2018, Else:2020jln}. Since the filling can be continuously tuned by applying a chemical potential, the defect parameter and thus $\gamma(\pi)$ can also be changed continuously. 

It also happens in some cases that the filling is fixed by additional global symmetries (such as charge conjugation), in which case $\gamma(\pi)$ becomes quantized. Typically, in these cases there is an exact LSM-type 't Hooft anomaly associated with the symmetries. For later references, we write down the expressions for $G(\theta)$ with $N$ unit cells and filling factor $\nu$:
\begin{equation}
    G(\theta)=
    \begin{cases}
        \euler^{\ii N\nu [\theta]_\pi }\euler^{-\alpha(\theta)N} & \theta\neq \pi\\
        2\cos(\pi N\nu)\euler^{-\alpha(\pi)N} & \theta=\pi
    \end{cases}.
    \label{cft-G-nu}
\end{equation}

The results obtained in this section rely on the assumption that the spectrum is described by a CFT. For lattice models, CFT only describes the spectrum up to a certain energy scale (e.g., of the order of the bandwidth). One may wonder whether the results still hold when the temperature is comparable (or even higher) to the cutoff scale. In the next sections, we turn to $G(\theta)$ in lattice models.

\section{Spin-1/2 chain with LSM anomaly}
We now consider a familiar system: the spin-$1/2$ chain with $\mathrm{O}(2)=\U\rtimes \Z_2$ internal symmetry. The $\U$ charge is given by
\begin{equation}
    Q=\frac12\sum_{n=1}^N \sigma^z_n,
\end{equation}
and the $\Z_2^C$ charge-conjugation symmetry is generated by 
\begin{equation}
  X=\prod_{n=1}^N \sigma^x_n.  
\end{equation}
Here $N$ is the number of sites. We will only consider spin chains with periodic boundary condition throughout this work.

Notice that the $\mathrm{O}(2)$ symmetry is on-site, so it is not anomalous on its own. However, because each spin-1/2 site carries a projective representation of the $\mathrm{O}(2)$ symmetry, the system exhibits a LSM anomaly between $\mathrm{O}(2)$ and lattice translation~\cite{ChengPRX2016, JianPRB2018, MetlitskiPRB2018, Else_LSM_2019}. 

In the absence of the $\Z_2$ symmetry, there is no 't Hooft anomaly between $\U$ and lattice translation. However, if the total filling is fixed, e.g., $Q=0$ (corresponding to half filling in the hard-core boson basis), the system exhibits the ``filling anomaly" for $\U$ and lattice translation~\cite{OshikawaLSM, ChengPRX2016, ChoPRB2017, JianPRB2018, Cheng:2022sgb}, forbidding the existence of a symmetric trivial state. Unlike the $\mathrm{O}(2)$ LSM anomaly, the filling anomaly only appears in the subspace constrained to have a fixed filling. Therefore, in a grand canonical ensemble, where the total charge is allowed to fluctuate, the filling anomaly is absent.

A prototypical model in this system is the XX Hamiltonian:
\begin{equation}
    H=-\sum_{n=1}^N (\sigma_n^x\sigma_{n+1}^x+\sigma_n^y\sigma_{n+1}^y).
\end{equation}
For this model, $G(\theta)$ can be computed with Jordan-Wigner transformation. We will sketch the key ingredients for the extraction of $\gamma(\theta)$, and leave the full details to Appendix~\ref{app:XX}. As it turns out, one needs to distinguish between the case of $\mu=0$ (half filling) and $\mu\neq 0$.

\subsection{\texorpdfstring{$\mu=0$: half filling}{μ=0: half filling}}\label{XX_mu=0}
First, we consider the model with full $\mathrm{O}(2)$ symmetry. By applying Jordan-Wigner transformation, the periodic XX Hamiltonian can be mapped to a free fermion Hamiltonian, which consists of two sectors. The first sector corresponds to an odd number of fermions and periodic boundary conditions for the fermions, while the second sector corresponds to an even number of fermions and antiperiodic boundary conditions for the fermions. The $\U$ charge $Q$ is mapped to the fermion number. Consequently, $G(\theta)$ can be expressed in terms of the following quantities:
\begin{equation}
    Z_{ss^{\prime}}(\theta,\beta) = \prod_{j=\frac{s^{\prime}}{2}}^{N-1+\frac{s^{\prime}}{2}}\kh{1+(-1)^s\euler^{{\ii \theta}}\euler^{4\beta \cos\frac{2\pi j}{N}}},\label{Zdef}
\end{equation}
where $s$ and $s^{\prime}$ take value in $0,1$. We have
\begin{multline}
    \Tr \euler^{\ii\theta Q}\euler^{-\beta H}=\\
    \frac12\fkh{Z_{00}(\theta,\beta)-Z_{10}(\theta,\beta)+Z_{01}(\theta,\beta)+Z_{11}(\theta,\beta)}.\label{XX_Eq_G}
\end{multline}
In particular, the partition function $\Tr\euler^{-\beta H}$ is obtained by setting $\theta=0$. And the first two terms $(Z_{00}(0,\beta)-Z_{10}(0,\beta))/2$ in Eq.~\eqref{XX_Eq_G} combine to be the partition function of free fermions with periodic boundary condition and odd number of fermions.

To evaluate $Z_{ss^{\prime}}(\theta,\beta)$, we use Euler-MacLaurin formula to convert $\ln Z_{ss^{\prime}}(\theta,\beta)$ into integrals. For example, for $\theta\neq\pi$:
\begin{equation}
    \ln Z_{00}(\theta,\beta) = \frac{N}{2\pi}I(\theta,\beta)+\bO(N^{-1}),
\end{equation}
where the integral $I(\theta,\beta)$ is defined as
\begin{align}
    I(\theta,\beta)&=\int_0^{2\pi}\ln\kh{1+\euler^{\ii{\theta}}\euler^{4\beta \cos x}}\diff x.
\end{align}
However, if some terms in the product Eq.~\eqref{Zdef} get close to 0, e.g., the term with $j$ near $N/4$ in $Z_{s=1,s'}(\theta=0,\beta)$, there are additional constant corrections. For example, we find that for $N\equiv 0\pmod{4}$:
\begin{align}
    \ln Z_{11}(0,\beta)&=\sum_{j=\frac{1}{2}}^{N-\frac{1}{2}} \ln\kh{1-\euler^{4\beta \cos\frac{2\pi j}{N}}}\notag\\
    &=\frac{N}{2\pi}\Re I(\pi,\beta)+\ln 4+\bO(N^{-1}).
\end{align}
The $\ln 4$ correction is crucial for the calculation of $\gamma(\pi)$.

Let us present the results. In the limit of large $N$, $G(\theta)$ takes the form given in Eq.~\eqref{cft-G-nu} with $\nu=1/2$. In particular, for $N$ even, we have
\begin{equation}
\begin{split}  
    \gamma(\theta)&=
        \begin{cases}
            0 & \theta\neq \pi\\
            \ln 2 & \theta=\pi
        \end{cases}.
\end{split}
\label{gamma-w-XX}
\end{equation}
It is worth noting that these results are valid for any temperature $\beta\neq 0$. For $\beta=0$, we instead get $G(\theta)=\kh{\cos\frac{\theta}{2}}^N$, which still gives $\gamma(\theta)=0$ for any $\theta \neq \pi$. However, since $G(\pi)=0$, $\gamma(\pi)$ is not well-defined.

To understand the physical meaning of $\gamma(\theta)$ and $\omega(\theta)$, let us consider the low-temperature limit $\beta\rightarrow \infty$. At low energy, the system is described by a $c=1$ Luttinger liquid (see, e.g., Ref.~\cite{Cheng:2022sgb}). The $\U$ charge $Q$ is identified with $Q_m$ in the low energy theory. The $\gamma(\pi)=\ln 2$ value for $N$ even can be easily understood from our general CFT result: in the presence of the $\euler^{\ii\pi Q}$ defect, corresponding to $\eta_m=1/2, \eta_w=0$, the zero mode Hamiltonian becomes
\begin{equation}
    H_0=Q_m^2+\frac14\kh{Q_w+\frac12}^2,
\end{equation}
and the ground states are twofold degenerate: $Q_m=0$ and $Q_w=0,-1$. This degeneracy is guaranteed by the LSM anomaly, as shown directly in the lattice model in Appendix~\ref{type-anomaly}. 

More generally, we need to first consider the translation symmetry. Importantly, the lattice unit translation leads to an emanant $\Z_2$ symmetry in the low-energy CFT. Namely, the lattice translation $T$ has the following representation in the low-energy theory:
\begin{equation}
    T = \euler^{-\ii\pi(Q_m+Q_w)} \euler^{\frac{2\pi \ii}{L}P},
\end{equation}
where $P$ is the CFT momentum. The lattice system with $N$ sites in the continuum limit becomes the CFT with $\eta_w=-N/2$, which then leads to Eq.~\eqref{cft-G-nu} with $\nu=1/2$.

While the results can be understood within the low-energy theory, we emphasize that our derivation in fact applies to any temperature $\beta\neq 0$, even at high temperature when the system is presumably not described by a CFT.

\begin{figure}[t]
    \centering
    \includegraphics[width=\columnwidth]{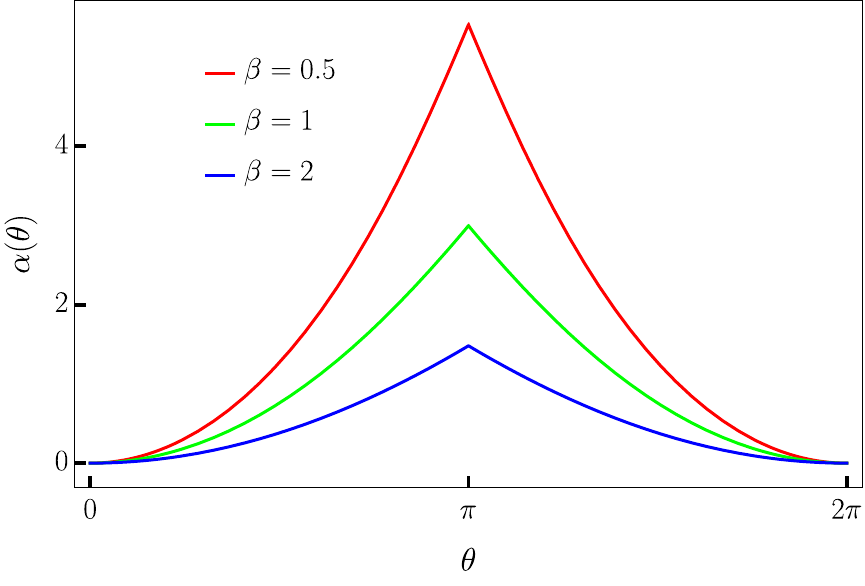}
    \caption{$\alpha(\theta)$ for $\nu=\frac{2}{3}$ at different $\beta$.}
    \label{fig_alpha}
\end{figure}

\begin{figure}[t]
  \centering
  \includegraphics[width=\columnwidth]{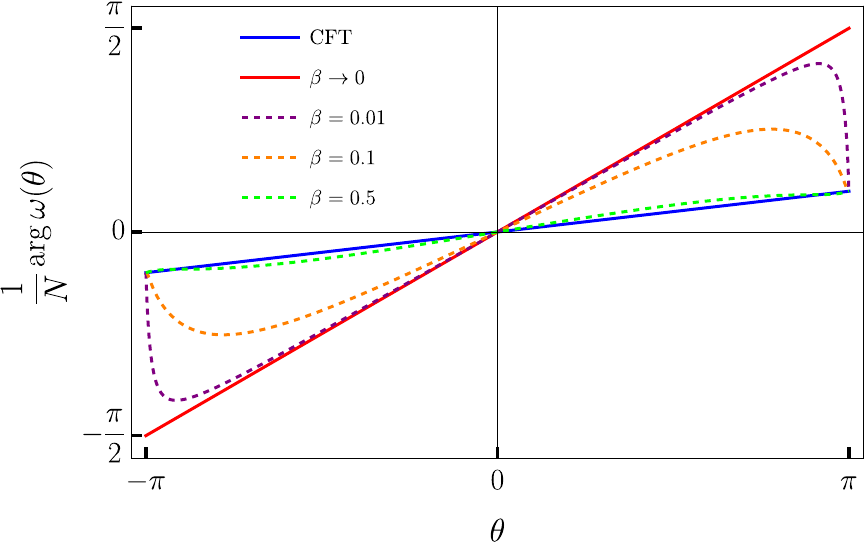}
  \caption{$\frac{1}{N}\arg\omega(\theta)$ for $\nu=1/10$. At low temperature, the numerical results agree with the CFT value, while at high temperture, they approach $\frac{1}{2}[\theta]_{\pi}$. For $\nu=1/2$, all the lines collapse.}
  \label{fig_phase}
\end{figure}

\subsection{\texorpdfstring{$\mu\neq0$}{μ not equals 0}}
It is instructive to consider turning on a nonzero chemical potential $\mu$ (or a nonzero Zeeman field in the spin language), which results in a different ground state filling factor (or magnetization plateau) in the XX model:
\begin{align}
    \nu=\frac{1}{\pi}\arccos\kh{-\frac{\mu}{4}}.
\end{align}
Note that we choose the convention that $\nu=1/2$ for $\mu=0$.

We find that for any finite $\beta\neq 0$ and $\theta\neq \pi\pmod{2\pi}$:
\begin{equation}
    \abs{G(\theta)}\approx\euler^{-\alpha(\theta)N}.
\end{equation}
From this, we observe that $\gamma(\theta)=0$. We plot $\alpha(\theta)$ for several different temperatures in Fig.~\ref{fig_alpha}. They agree very well with the CFT result, and exhibit a cusp at $\theta=\pi$ for all temperatures. The cusp becomes sharper as the temperature increases.

The phase factor $\omega(\theta)$ now depends on $\beta$ and $\nu$ in a complicated way. This is expected, since for small $\theta$, one has
\begin{equation}
    \omega(\theta)\approx \euler^{\ii N \nu(\beta)\theta},
\end{equation}
where $\nu(\beta)$ is the average filling at temperature $\beta$.
Only when $\beta$ is large (i.e. low temperature), the phase factor approaches the CFT form $\euler^{\ii N\nu[\theta]_\pi}$. We plot $\arg(\omega)$ for the $\nu=1/10$ case in Fig~\ref{fig_phase}.

For $\theta=\pi$, we are able to determine the phase factor as well:
\begin{equation}
    G(\pi)\approx 2\cos(\pi N\nu)\euler^{-\alpha(\pi)N}.\label{Gpi-2}
\end{equation}
Notably, these results hold for any finite temperature, as long as $\beta>0$.

From the perspective of low-energy theory, the emanant symmetry from translation becomes $\euler^{-2\pi \ii\nu Q_w}$. A lattice system of $N$ sites flows to a $\U$ free boson CFT with a defect $\eta_w=-N\nu$. Eq.~\eqref{Gpi-2} then immediately follows from Eq.~\eqref{gamma-with-defect}.

This example suggests that $\gamma(\pi)$ is not quantized in general in a lattice system, as it varies smoothly with the chemical potential (or filling factor). However, in the presence of additional symmetry, such as charge conjugation symmetry that fixes the filling, $\gamma(\pi)$ takes quantized value $\ln 2$ (for even $N$). A spin-1/2 chain with $\mathrm{O}(2)$ symmetry has the LSM anomaly, which is a true 't Hooft anomaly that holds for the entire Hilbert space. For this reason we expect that the quantization of $\gamma(\pi)$ holds more generally. We show in the next subsection that this is indeed the case, and $\gamma(\pi)=\ln 2$ for even $N$ is a direct consequence of the LSM anomaly.

\subsection{\texorpdfstring{Quantization of $\gamma(\pi)$}{Quantization of γ(π)}}\label{O2-quant}
We now show that $\gamma(\pi)=\ln 2$ holds for any thermal state of a local Hamiltonian (i.e. with short-range interactions), as long as the $\mathrm{O}(2)=\U\rtimes\Z_2^C$ symmetry is preserved. In fact, as will become clear, all we need is the $\Z_2\times\Z_2$ subgroup, generated by $Z=\prod_n \sigma^z_n$ and $X=\prod_n \sigma^x_n$. We define
\begin{equation}
    G_Z=\Tr \rho Z.
\end{equation}
Note that $G(\pi)=\ii^N G_Z$.

To proceed we follow the argument presented in Ref.~\cite{Zang}. First we represent a translation-invariant density matrix as a matrix product operator (MPO)~\cite{Verstraete2004,perezgarcia2007matrix, Cirac2017, Cirac2021matrix}:
\begin{equation}
\begin{aligned}
    \rho&=\sum_{\{s\},\{s'\}}\Tr\left[ \cdots \hat{M}^{s_{j}s'_{j}}\hat{M}^{s_{j+1}s'_{j+1}}\cdots \right]\\
    & \qquad \qquad \times\ket{\dots s_j s_{j+1},\dots}\bra{\dots s'_j s'_{j+1} \dots}\\
    & =\quad \begin{tikzpicture}[scale=1,baseline={([yshift=-2pt]current bounding box.center)}]
    \draw[white] (0pt,-20pt) -- (0pt,20pt);
    \draw (-8pt,0pt) -- (46pt,0pt);
    \draw[densely dashed] (46pt,0pt) -- (78pt,0pt);
    \draw (78pt,0pt) -- (104pt,0pt);
    \draw (-8pt,0pt) arc (90:270:2.5pt);
    \draw (104pt,0pt) arc (90:-90:2.5pt);
    \draw [densely dotted] (-8pt,-5pt) -- (-2pt,-5pt);
    \draw [densely dotted] (104pt,-5pt) -- (98pt,-5pt);
    \draw (8pt,-16pt) -- (8pt,16pt);
    \draw (36pt,-16pt) -- (36pt,16pt);
    \draw (88pt,-16pt) -- (88pt,16pt);
    \filldraw[thick,fill=white] (0pt,-8pt) rectangle (16pt,8pt);
    \node at (8pt,0pt) {$M$};
    \filldraw[thick,fill=white] (28pt,-8pt) rectangle (44pt,8pt);
    \node at (36pt,0pt) {$M$};
    \filldraw[thick,fill=white] (80pt,-8pt) rectangle (96pt,8pt);
    \node at (88pt,0pt) {$M$};
    \end{tikzpicture}\quad .
    \label{MPO-rho}
\end{aligned}
\end{equation}
It is known that such an approximation is always possible for thermal states of local Hamiltonians~\cite{HastingsPRB2006, MolnarPRB2015, KuwaharaPRX2021, Huang2021}. Furthermore, since there is no long-range order at finite temperature in 1D, we expect that the MPO is injective~\cite{perezgarcia2007matrix, Cirac2021matrix}, i.e., the corresponding transfer matrix has a nondegenerate leading eigenvalue.

Using the MPO representation, we write
\begin{equation}
    \begin{aligned}
    G_Z=\Tr \rho\prod_n\sigma^z_n
    &=
    \begin{tikzpicture}[scale=1,baseline={([yshift=-2pt]current bounding box.center)}]
    \draw (8pt,-20pt) -- (8pt,20pt);
    \draw (8pt,20pt) arc (180:0:2.5pt);
    \draw (8pt,-20pt) arc (180:360:2.5pt);
    \draw [densely dotted] (13pt,20pt) -- (13pt,8pt);
    \draw [densely dotted] (13pt,-20pt) -- (13pt,-8pt);
    \draw (36pt,-20pt) -- (36pt,20pt);
    \draw (36pt,20pt) arc (180:0:2.5pt);
    \draw (36pt,-20pt) arc (180:360:2.5pt);
    \draw [densely dotted] (41pt,20pt) -- (41pt,8pt);
    \draw [densely dotted] (41pt,-20pt) -- (41pt,-8pt);
    \draw (88pt,-20pt) -- (88pt,20pt);
    \draw (88pt,20pt) arc (180:0:2.5pt);
    \draw (88pt,-20pt) arc (180:360:2.5pt);
    \draw [densely dotted] (93pt,20pt) -- (93pt,8pt);
    \draw [densely dotted] (93pt,-20pt) -- (93pt,-8pt);
    \draw[white] (0pt,-20pt) -- (0pt,20pt);
    \draw (-8pt,0pt) -- (46pt,0pt);
    \draw[densely dashed] (46pt,0pt) -- (78pt,0pt);
    \draw (78pt,0pt) -- (104pt,0pt);
    \draw (-8pt,0pt) arc (90:270:2.5pt);
    \draw (104pt,0pt) arc (90:-90:2.5pt);
    \draw [densely dotted] (-8pt,-5pt) -- (-2pt,-5pt);
    \draw [densely dotted] (104pt,-5pt) -- (98pt,-5pt);
    \draw (8pt,-16pt) -- (8pt,16pt);
    \draw (36pt,-16pt) -- (36pt,16pt);
    \draw (88pt,-16pt) -- (88pt,16pt);
    \filldraw[thick,fill=white] (0pt,-8pt) rectangle (16pt,8pt);
    \node at (8pt,0pt) {$M$};
    \filldraw[thick,fill=white] (28pt,-8pt) rectangle (44pt,8pt);
    \node at (36pt,0pt) {$M$};
    \filldraw[thick,fill=white] (80pt,-8pt) rectangle (96pt,8pt);
    \node at (88pt,0pt) {$M$};
    \filldraw [red,fill=red] (88pt,15pt) circle (1pt) node[left] {$\sigma^z$};
    \filldraw [red,fill=red] (8pt,15pt) circle (1pt) node[left] {$\sigma^z$};
    \filldraw [red,fill=red] (36pt,15pt) circle (1pt) node[left] {$\sigma^z$};
    \end{tikzpicture} \\
    &\eqcolon\Tr\fkh{(M_Z)^N}. 
    \end{aligned}
\end{equation}
Here we define the ``symmetry-twisted" transfer matrix ${M}_Z$ as 
\begin{equation}
    M_Z \coloneq
    \begin{tikzpicture}[scale=1,baseline={([yshift=-2pt]current bounding box.center)}]
    \draw (8pt,-20pt) -- (8pt,20pt);
    \draw (-8pt,0pt) -- (24pt,0pt);
    \draw (8pt,20pt) arc (180:0:2.5pt);
    \draw (8pt,-20pt) arc (180:360:2.5pt);
    \draw [densely dotted] (13pt,20pt) -- (13pt,8pt);
    \draw [densely dotted] (13pt,-20pt) -- (13pt,-8pt);
    \filldraw[thick,fill=white] (0pt,-8pt) rectangle (16pt,8pt);
    \node at (8pt,0pt) {$M$};
    \filldraw [red,fill=red] (8pt,15pt) circle (1pt) node[left] {$\sigma^z$};
    \end{tikzpicture} 
    ~~=\sum_{ss'}\sigma^z_{s's}\cdot {M}^{ss'}.
\end{equation}

Crucially, the density matrix also commutes with the $X$ symmetry: $X\rho X=\rho$. Via the fundamental theorem of matrix product vectors, there must exist an invertible matrix $W_X$ such that
\begin{equation}
    \begin{tikzpicture}[scale=1,baseline={([yshift=-2pt]current bounding box.center)}]
    \draw (8pt,-20pt) -- (8pt,20pt);
    \draw (-12pt,0pt) -- (28pt,0pt);
    \filldraw[thick,fill=white] (0pt,-8pt) rectangle (16pt,8pt);
    \node at (8pt,0pt) {$M$};
    \end{tikzpicture}=
    \begin{tikzpicture}[scale=1,baseline={([yshift=-2pt]current bounding box.center)}]
    \draw (8pt,-20pt) -- (8pt,20pt);
    \draw (-12pt,0pt) -- (28pt,0pt);
    \filldraw[thick,fill=white] (0pt,-8pt) rectangle (16pt,8pt);
    \node at (8pt,0pt) {$M$};
    \filldraw [red,fill=red] (8pt,15pt) circle (1pt) node[left] {$\sigma^x$};
    \filldraw [red,fill=red] (8pt,-15pt) circle (1pt) node[left] {$\sigma^x$};
    \filldraw [red,fill=red] (-9pt,0pt) circle (1pt) node[left] {$W_X$};
    \filldraw [red,fill=red] (25pt,0pt) circle (1pt) node[right] {$W_X^{-1}$};
    \end{tikzpicture}.
    \label{X-symmetry-1}
\end{equation}
Applying this virtual symmetry to ${M}_Z$, it follows from the relation $\sigma^x \sigma^z \sigma^x=-\sigma^z$ that
\begin{equation}
    W_X M_Z W_X^{-1} = -M_Z.\label{mzx}
\end{equation}
In other words, $W_X$ and ${M}_Z$ anticommute. Therefore all eigenvalues of ${M}_Z$ must come in pairs $\pm\lambda$, including those with the largest magnitude. In principle, it is possible that $\abs{\lambda}$ has degeneracy more than 2. However, given that the leading eigenvalue of $M$ is nondegenerate, we expect that $M_Z$ has no more degeneracy in the leading $|\lambda|$ than what is required by the symmetry condition in Eq.~\eqref{mzx}. That is, the leading eigenvalues of ${M}_Z$ should be $\pm \lambda_{\rm max}$~\footnote{It is possible that $\lambda_{\rm max}=0$, in which case $\gamma$ becomes meaningless. This is what happens at $\beta=0$.}. With this assumption,
\begin{equation}
    G_Z\approx\fkh{1+(-1)^N}\lambda_{\rm max}^N=\begin{cases}
    2\lambda_{\rm max}^N & N\text{ is even}\\
    0 & N\text{ is odd}
\end{cases}.
\end{equation}
From this result we immediately see that $\gamma(\pi)=\ln 2$.

Notice that the result applies to any system with $\Z_2\times\Z_2$ LSM anomaly~\cite{ChengPRX2016, PoPRL2017,  ogata2019lieb, OgataLSM}, since the only symmetry property used here is $\sigma^x \sigma^z=-\sigma^z \sigma^x$. In fact, the only assumptions needed on $\rho$ are translation invariance and the $X$ symmetry.

Naturally one may wonder whether a similar argument can explain the $(-1)^{N/2}$ factor in $G(\pi)$ of the XX spin chain. We now make a further assumption that $M$ is Hermitian (within a certain gauge), and $\sigma^z$ is real (i.e. has real matrix elements). In this case, we can easily show that $M_Z$ is also Hermitian:
\begin{align}
    \fkh{M_Z}_{ba} &= \sum_{ss'}\sigma^z_{s's}{M}^{ss'}_{ba}\\
    &=\sum_{ss'}\sigma^z_{s's}\kh{{M}^{ss'}_{ab}}^*\\
    &=\kh{\sum_{ss'}\sigma^z_{s's}{M}^{ss'}_{ab}}^*\\
    &=\fkh{M_Z}_{ab}^*.
\end{align}

As a result, the eigenvalues of ${M}_Z$ come in pairs $\pm|\lambda|$, and for even $N$ we have
\begin{equation}
    G_Z\approx 2\abs{\lambda_{\rm max}}^N.\label{Gz}
\end{equation}

While generally a transfer matrix may not be Hermitian, it is known that, e.g., the transfer matrix of the XXZ model
\begin{equation}
    H_{\rm XXZ}=\sum_{n=1}^N (\sigma_n^x\sigma_{n+1}^x+\sigma_n^y\sigma_{n+1}^y+\Delta \sigma^z_n\sigma^z_{n+1})\label{XXZ}
\end{equation}
can be made Hermitian for $\Delta\geq 0$~\cite{TangPRL2020}. Therefore Eq.~\eqref{Gz} applies to $H_{\rm XXZ}$ as well.

We note that the translation invariance is crucial for this argument, otherwise Eq.~\eqref{MPO-rho} would not hold. In fact, if the translation symmetry is broken (e.g., the unit cell is doubled) but the $\mathrm{O}(2)$ symmetry is preserved, one can easily construct examples with $\gamma(\theta)=0$ for all $\theta$.

\section{\texorpdfstring{Spin chain with mixed $\mathrm{O}(2)\times\Z_2$ anomaly}{Spin chain with mixed O(2)xZ2 anomaly}}
In this section, we consider a spin chain with anomalous $\mathrm{O}(2)\times \Z_2^X$ symmetry. Unlike the previous example, here the $\U$ symmetry is not on-site, and importantly there is a type-III 't Hooft anomaly between $\mathrm{O}(2)$ and $\Z_2^X$.

More specifically, the system is made of qubits, with the $\U$ charge defined as
\begin{align}
    Q=\frac{1}{4}\sum_{n=1}^N (1-\sigma_n^z\sigma_{n+1}^z).
\end{align}
In the $\sigma^z$ eigenbasis, $2Q$ counts the number of domain walls. With periodic boundary conditions, $Q$ takes integer values. Interestingly, $Q$ is manifestly not on-site, but still anomaly-free. The constant term (equals to $N/4$) is included in $Q$ to ensure that $Q$ takes integer values for any system size $N$.

The charge-conjugation symmetry $\Z_2^C$ in the $\mathrm{O}(2)$ group is generated by 
\begin{equation}
    X_{\rm even}=\prod_{n=1}^{N/2}\sigma^x_{2n}.
\end{equation}
One can readily see that
\begin{equation}
    X_{\rm even}QX_{\rm even}^{-1}= \frac{1}{4}\sum_{n=1}^N (1+\sigma_n^z\sigma_{n+1}^z)=-Q+\frac{N}{2}.\label{C-on-Q}
\end{equation}
Thus, to preserve the $\mathrm{O}(2)$ symmetry, the filling factor $\nu=Q/N$ must be $1/4$. We can similarly define $X_{\rm odd}$, which has the same action on $Q$.

Lastly, the on-site $\Z_2^X$ symmetry is generated by 
\begin{equation}
  X=\prod_{n=1}^N \sigma_n^x=X_{\rm even}X_{\rm odd}.
\end{equation}

Now we discuss the 't Hooft anomaly of the internal $\mathrm{O}(2)\times\Z_2^X$ symmetry. As mentioned earlier, there is a mixed anomaly between $\mathrm{O}(2)$ and $\Z_2^X$. In fact, the same kind of anomaly is already present when we restrict to $\Z_2\subset \U$. The $\Z_2\times \Z_2^{C}\times\Z_2^X$ symmetry has an anomaly associated with the so-called ``type-III" 3-cocycle~\cite{deWildPropitius:1995cf}. Physically, it is characterized by the $\Z_2$ defect transforming projectively under $\Z_2^{C}\times\Z_2^X$. 

Moreover, even if the $\Z_2^C$ symmetry is ignored, the remaining $\U\times\Z_2^X$ symmetry is still anomalous. The system can be viewed as a lattice model for the edge of a bosonic SPT state protected by the $\U\times\Z_2$ symmetry~\cite{Chen_2013, metlitski20191d, Son_2019, horinouchi2020solvable}. Turning on a nonzero chemical potential breaks $\Z_2^C$ but preserves $\Z_2^X$.

Below we will study the following Levin-Gu Hamiltonian preserving $\mathrm{O}(2)\times\Z_2^X$:
\begin{align}
    H_{\rm LG}=-\sum_{n=1}^N (\sigma_n^x-\sigma_{n-1}^z\sigma_n^x\sigma_{n+1}^z),
\end{align}
which was first considered in \cite{LevinGu} as an edge model of the nontrivial (2+1)d $\Z_2$ SPT state. The model is exactly solvable, and can be mapped to a gauged XX spin chain as follows. Introduce a dual representation of the system~\cite{LevinGu}:
\begin{align}
    \begin{aligned}
        \sigma_n^x&=\tau_{n-1}^x\tau_n^x\mu_{n-1,n}^z,\\
        \sigma_n^y&=-\tau_{n-1}^x\tau_n^x\mu_{n-1,n}^y,\\
        \sigma_n^z&=\mu_{n-1,n}^x,
    \end{aligned}
\end{align}
where $\tau$ can be understood as domain wall variables and $\mu$ represent $\Z_2$ gauge fields. They are subject to the Gauss's law constraint:
\begin{equation}
   \tau_n^z=\mu_{n-1,n}^x\mu_{n,n+1}^x. 
\end{equation}
In this representation, $Q$ can be written as
\begin{align}
    Q=\frac{1}{4}\sum_n(1-\tau_n^z).
\end{align}
Similarly, $X=\prod_n \mu_{n,n+1}^x$. The Levin-Gu Hamiltonian becomes
\begin{equation}
    H_{\rm LG}=-\sum_{n=1}^N (\tau_{n}^x\tau_{n+1}^x+\tau_{n}^y\tau_{n+1}^y)\mu_{n,n+1}^z,
\end{equation}
resembling a XX spin chain coupled to a $\Z_2$ gauge field. $Q$ is the total spin-$z$ component and $X$ is the Wilson loop for the gauge field.  

Similar to Sec.~\ref{XX_mu=0}, $G(\theta)$ can be computed by mapping to free fermions. We find that for any $\beta\geq 0$, $G(\theta)$ is given precisely by Eq.~\eqref{cft-G-nu} with $\nu=1/4$. In particular, $\gamma(\theta)$ is given by
\begin{equation}
\begin{split}
   \gamma(\theta)&=0, \quad \theta\neq \pi\pmod{2\pi},\\
     \gamma(\pi)&=\ln \abs{2\cos\frac{N\pi}{4}}\\
     &=\begin{cases}
        \ln 2 & N\equiv 0 \pmod{4}\\
        \ln \sqrt{2} & N\equiv \pm 1\pmod{4}
    \end{cases}. 
\end{split}\label{gamma-LG}      
\end{equation}
Details of the calculations can be found in Appendix~\ref{app:LG}.

When $N\equiv 2\pmod{4}$, one can show on general grounds that $G(\pi)=0$: As $N\equiv 2\pmod{4}$, $N/2$ is an odd integer. It then follows that
\begin{align}
    \Tr \euler^{\ii\pi Q}\euler^{-\beta H} &= \Tr X_{\rm even}^2\euler^{\ii\pi Q} \euler^{-\beta H}\\
    &=(-1)^{N/2}\Tr \euler^{-\ii\pi Q}\euler^{-\beta H}\\
    &=-\Tr \euler^{\ii\pi Q}\euler^{-\beta H}.
\end{align}
Thus we conclude that $G(\pi)=0$. Essentially, the system forms a nontrivial projective representation of $\mathrm{O}(2)$.

To understand the physical meaning of $\gamma(\theta)$ and $\omega(\theta)$, let us consider two limits. First, at low temperature $\beta\rightarrow \infty$, the system is described by a $c=1$ Luttinger liquid with $R=\frac{1}{\sqrt{2}}$~\cite{Cheng:2022sgb}. The $\U$ charge $Q$ is identified with $Q_m$ in the low energy theory, and $X$ becomes $\euler^{\ii\pi Q_w}$. In addition, the lattice unit translation leads to an emanant $\Z_4$ symmetry $\euler^{-\frac{\ii\pi}{2}Q_w}$~\cite{Cheng:2022sgb}. Namely, the lattice translation $T$ has the following representation in the low-energy theory:
\begin{equation}
    T = \euler^{-\frac{\ii\pi}{2}Q_w} \euler^{\frac{2\pi\ii}{L}P},
\end{equation}
where $P$ is the CFT momentum.  As a result of the nontrivial emanant symmetry, a chain of size $N$ should flow to a CFT with a $\eta_w=-N/4$ defect. Therefore, in this limit $G(\theta)$ takes the form of Eq.~\eqref{cft-G-nu} with $\nu=1/4$.

When $N\equiv 0\pmod{4}$, $\gamma(\pi)=\ln 2$ is also expected from the mixed anomaly between $\U$ and $\Z_2$, as there must be level crossing between states with opposite $\Z_2$ quantum numbers. The degeneracy at that point must be at least $2$. With the $\Z_2^C$ symmetry, the crossing must happen at $\theta=\pi$. Alternatively, the degeneracy at $\theta=\pi$ follows from the type-III anomaly for the $\Z_2\times\Z_2^{C}\times\Z_2^X$ symmetry as shown explicitly in Appendix~\ref{type-anomaly}.

We note in passing that adding a chemical potential [thus reducing $\mathrm{O}(2)$ to $\U$] changes the ground state filling factor, and similar to the case of the XX model, the value of $\gamma(\pi)$ also changes continuously with the chemical potential.

The other limit is $\beta\rightarrow 0$, i.e., high temperature. In this case, we can neglect the Hamiltonian and compute the trace of the operator $\euler^{\ii\theta Q}$. This computation will be discussed in the next subsection.

In both limits, one finds that $\gamma(\pi)$ takes the quantized values given in Eq.~\eqref{gamma-LG} as long as the $\Z_2^C\times\Z_2^X$ symmetry is preserved, which suggests that the quantization is independent of the details of the Hamiltonian. This will be established in Sec.~\ref{quant-gamma-LG}.

\subsection{High temperature limit}\label{LG_HighT}
It is instructive to consider the infinite-temperature state $\rho=\frac{1}{2^N}\id$ first. We have
\begin{align}
    {G(\theta)}&=2^{-N}{\Tr\euler^{\ii\theta Q}}\\
    &=2^{-N}\euler^{\frac{\ii\theta}{4}N}\Tr \euler^{-\ii\frac{\theta}{4}\sum_n \sigma_n^z\sigma_{n+1}^z}.
\end{align}
Note that the trace is precisely calculating partition function of 1D Ising Hamiltonian with imaginary coupling. Generally, we can write
\begin{align}
    \Tr \exp\kh{K\sum_{n=1}^N\sigma_n^z\sigma_{n+1}^z}=z_+^N+z_-^N,
\end{align}
where $z_\pm=\euler^K\pm\euler^{-K}$ are eigenvalues of the transfer matrix. In our case $K=-\frac{1}{4}\ii\theta$, so we have
\begin{align}
    2^{-N}\Tr \euler^{-\frac14\ii\theta\sum_n \sigma_n^z\sigma_{n+1}^z} = \cos^N\frac{\theta}{4}+(-\ii)^N\sin^N\frac{\theta}{4}.
\end{align}

First, let us look at the special case $\theta=\pi$. We have
\begin{align}
G(\pi)&=\euler^{\frac14\ii \pi N}[1+(-\ii)^{N}]2^{-N/2}\\
&=2\cos \kh{\frac{\pi N}{4}} 2^{-N/2}.
\end{align}

For general $\theta\in(0,2\pi)$,
\begin{align}
    \ln\abs{G(\theta)} =\ln\abs{\cos^N\frac{\theta}{4}+(-\ii)^N\sin^N\frac{\theta}{4}}.
\end{align}
If $\theta<\pi$, then $\cos\frac{\theta}{4}>\sin\frac{\theta}{4}$, $\ln\abs{G(\theta)}\approx N\ln\cos\frac{\theta}{4}$ for $N$ large. If $\pi<\theta<2\pi$, then $\sin\frac{\theta}{4}>\cos\frac{\theta}{4}$, $\ln\abs{G(\theta)}\approx N\ln\sin\frac{\theta}{4}$ for $N$ large. In both cases, we have $\gamma(\theta)=0$ as claimed. 

We now consider a slightly deformed state, adding a chemical potential: $\rho\propto \euler^{\lambda Q}$. Notice that $\lambda\neq 0$ breaks the $\Z_2^C$ symmetry, but the $\Z_2^X$ symmetry is preserved. It is straightforward to generalize the calculations above, and here we will just present the $\theta=\pi$ result:
\begin{equation}
    G(\pi)=2^{-\frac{N}{2}} \fkh{(1-\ii \tanh\lambda)^N + (-\ii)^N (1+\ii \tanh \lambda)^N}.
\end{equation}
We can extract 
\begin{align}
    \gamma(\pi)=\ln \abs{2\cos N\kh{\varphi-\frac{\pi}{4}}},
\end{align}
where $\varphi(\lambda)=\arctan(\tanh\lambda)$. One can also easily show that $\gamma(\theta)=0$ for $\theta\neq \pi$. Notice that in this state $\rho$, the $\U$ charge density is given by $\frac{1}{4}(1-\tanh \lambda)$. This simple example shows that $\gamma(\pi)$ changes with the filling and is not quantized with just $\U\times\Z_2^X$.

\subsection{\texorpdfstring{Quantization of $\gamma(\pi)$}{Quantization of γ(π)}}\label{quant-gamma-LG}
In this section, we present arguments for the quantization of $\gamma(\pi)$ in this system. Before going to the details, it is important to clarify the role of translation invariance. According to Eq.~\eqref{C-on-Q}, with full translation symmetry, the filling factor of the system is fixed at $1/4$, which is already a strong hint that $\gamma(\pi)$ should be universal. Even with a doubled unit cell, the filling factor is $1/2$ and one would expect that $\gamma(\pi)$ is quantized to $\ln 2$ just like the $\mathrm{O}(2)$ LSM case discussed in Sec.~\ref{O2-quant} (even though the $\U$ charge is not on-site, an important difference). However, we will find that the quantization $\gamma(\pi)=\ln 2$ does not really rely on LSM-type anomaly and is instead enforced by the anomalous internal symmetry. More specifically, we will show that the quantization holds with a four-site unit cell for which the filling factor is an integer. On the other hand, if the state has the full translation symmetry, $\gamma(\pi)$ exhibits interesting dependence on $N\pmod{4}$. We will present general arguments to explain this behavior.

First, because the symmetry operator $\euler^{\ii\pi Q}$ is not on-site, it is represented as a MPO with the following tensor with bond dimension $D=2$ (up to an overall phase):
\begin{equation}
    \begin{tikzpicture}[scale=1,baseline={([yshift=-2pt]current bounding box.center)}]
    \draw (8pt,-20pt) node [below] {$s$} -- (8pt,20pt) node [above] {$s'$};
    \draw (-12pt,0pt) node [left] {$\alpha$} -- (28pt,0pt) node[right] {$\beta$};
    \filldraw[thick,fill=white] (0pt,-8pt) rectangle (16pt,8pt);
    \node at (8pt,0pt) {$U$};
    \end{tikzpicture}=\delta_{\alpha s}\ii^{\alpha+\beta}(-1)^{\alpha\beta}.
\end{equation}
Here $\alpha, \beta=0,1$ are the bond indices.

It is straightforward to check that the tensor satisfies the following two conditions:
\begin{align}
    \begin{tikzpicture}[scale=1,baseline={([yshift=-2pt]current bounding box.center)}]
    \draw (8pt,-20pt) -- (8pt,20pt);
    \draw (-12pt,0pt) -- (28pt,0pt);
    \filldraw[thick,fill=white] (0pt,-8pt) rectangle (16pt,8pt);
    \node at (8pt,0pt) {$U$};
    \end{tikzpicture}&=
    \begin{tikzpicture}[scale=1,baseline={([yshift=-2pt]current bounding box.center)}]
    \draw (8pt,-20pt) -- (8pt,20pt);
    \draw (-12pt,0pt) -- (28pt,0pt);
    \filldraw[thick,fill=white] (0pt,-8pt) rectangle (16pt,8pt);
    \node at (8pt,0pt) {$U$};
    \filldraw [red,fill=red] (8pt,15pt) circle (1pt) node[left] {$\sigma^x$};
    \filldraw [red,fill=red] (8pt,-15pt) circle (1pt) node[left] {$\sigma^x$};
    \filldraw [red,fill=red] (-9pt,0pt) circle (1pt) node[left] {$\sigma^x$};
    \filldraw [red,fill=red] (25pt,0pt) circle (1pt) node[right] {$\sigma^x$};
    \end{tikzpicture}\\ 
    \begin{tikzpicture}[scale=1,baseline={([yshift=-2pt]current bounding box.center)}]
    \draw (8pt,-20pt) -- (8pt,20pt);
    \draw (-12pt,0pt) -- (28pt,0pt);
    \filldraw[thick,fill=white] (0pt,-8pt) rectangle (16pt,8pt);
    \node at (8pt,0pt) {$U$};
    \end{tikzpicture}&=-
    \begin{tikzpicture}[scale=1,baseline={([yshift=-2pt]current bounding box.center)}]
    \draw (8pt,-20pt) -- (8pt,20pt);
    \draw (-12pt,0pt) -- (28pt,0pt);
    \filldraw[thick,fill=white] (0pt,-8pt) rectangle (16pt,8pt);
    \node at (8pt,0pt) {$U$};
    \filldraw [red,fill=red] (-9pt,0pt) circle (1pt) node[left] {$\sigma^z$};
    \filldraw [red,fill=red] (25pt,0pt) circle (1pt) node[right] {$\sigma^y$};
    \end{tikzpicture} .
\end{align}
The first condition guarantees that $\euler^{\ii\pi Q}$ commutes with $X$. The second condition can be understood as a kind of ``gauge symmetry" of the $U$ tensor.

$G(\pi)$ can be written as 
\begin{equation}
    G(\pi)=\Tr M_U^N,
\end{equation}
where the tensor $M_U$ is defined by the following diagram:
\begin{equation}
    M_U \coloneq
    \begin{tikzpicture}[scale=1,baseline={([yshift=-2pt]current bounding box.center)}]
    \draw (8pt,-20pt) -- (8pt,40pt);
    \draw (-8pt,0pt) -- (24pt,0pt);
    \draw (-8pt,22pt) -- (24pt,22pt);
    \draw (8pt,40pt) arc (180:0:2.5pt);
    \draw (8pt,-20pt) arc (180:360:2.5pt);
    \draw [densely dotted] (13pt,40pt) -- (13pt,8pt);
    \draw [densely dotted] (13pt,-20pt) -- (13pt,-8pt);
    \filldraw[thick,fill=white] (0pt,14pt) rectangle (16pt,30pt);
    \node at (8pt,22pt) {$U$};
    \filldraw[thick,fill=white] (0pt,-8pt) rectangle (16pt,8pt);
    \node at (8pt,0pt) {$M$};
    \end{tikzpicture}. 
\end{equation}

Now we consider the $\Z_2^C$ symmetry. To this end, it is convenient to group two neighboring sites into a doubled unit cell. By our assumption, the density matrix commutes with $X_{\rm even}$, and we must have
\begin{align}
   \begin{tikzpicture}[scale=1,baseline={([yshift=-2pt]current bounding box.center)}]
    \draw (8pt,-20pt) -- (8pt,20pt);
    \draw (32pt,-20pt) -- (32pt,20pt);
    \draw (-12pt,0pt) -- (52pt,0pt);
    \filldraw[thick,fill=white] (0pt,-8pt) rectangle (16pt,8pt);
    \node at (8pt,0pt) {$M$};
    \filldraw[thick,fill=white] (24pt,-8pt) rectangle (40pt,8pt);
    \node at (32pt,0pt) {$M$};
    \end{tikzpicture}= 
    \begin{tikzpicture}[scale=1,baseline={([yshift=-2pt]current bounding box.center)}]
    \draw (8pt,-20pt) -- (8pt,20pt);
    \draw (32pt,-20pt) -- (32pt,20pt);
    \draw (-12pt,0pt) -- (52pt,0pt);
    \filldraw[thick,fill=white] (0pt,-8pt) rectangle (16pt,8pt);
    \node at (8pt,0pt) {$M$};
    \filldraw[thick,fill=white] (24pt,-8pt) rectangle (40pt,8pt);
    \node at (32pt,0pt) {$M$};
    \filldraw [red,fill=red] (32pt,15pt) circle (1pt) node[left] {$\sigma^x$};
    \filldraw [red,fill=red] (32pt,-15pt) circle (1pt) node[left] {$\sigma^x$};
    \filldraw [red,fill=red] (-9pt,0pt) circle (1pt) node[left] {$V_X$};
    \filldraw [red,fill=red] (49pt,0pt) circle (1pt) node[right] {$V_X^{-1}$};
    \end{tikzpicture}.\label{Rsym-MPO}
\end{align}
Here $V_X$ is an invertible matrix.

We now prove the following key relation:
\begin{equation}
    (\sigma^z \otimes V_X)M_U^2=-M_U^2 (\sigma^z \otimes V_X).
\end{equation}
It can be established by the following steps:
\begin{align}
   \begin{tikzpicture}[scale=1,baseline={([yshift=-2pt]current bounding box.center)}]
    \draw (8pt,-20pt) -- (8pt,44pt);
    \draw (-8pt,0pt) -- (52pt,0pt);
    \draw (-8pt,26pt) -- (52pt,26pt);
    \draw (8pt,44pt) arc (180:0:2.5pt);
    \draw (8pt,-20pt) arc (180:360:2.5pt);
    \draw [densely dotted] (13pt,44pt) -- (13pt,8pt);
    \draw [densely dotted] (13pt,-20pt) -- (13pt,-8pt);
    \draw (36pt,-20pt) -- (36pt,44pt);
    \draw (36pt,44pt) arc (180:0:2.5pt);
    \draw (36pt,-20pt) arc (180:360:2.5pt);
    \draw [densely dotted] (41pt,44pt) -- (41pt,8pt);
    \draw [densely dotted] (41pt,-20pt) -- (41pt,-8pt);
    \filldraw[thick,fill=white] (0pt,18pt) rectangle (16pt,34pt);
    \node at (8pt,26pt) {$U$};
    \filldraw[thick,fill=white] (0pt,-8pt) rectangle (16pt,8pt);
    \node at (8pt,0pt) {$M$};
    \filldraw[thick,fill=white] (28pt,18pt) rectangle (44pt,34pt);
    \node at (36pt,26pt) {$U$};
    \filldraw[thick,fill=white] (28pt,-8pt) rectangle (44pt,8pt);
    \node at (36pt,0pt) {$M$};
    \end{tikzpicture}
    &=\ii
    \begin{tikzpicture}[scale=1,baseline={([yshift=-2pt]current bounding box.center)}]
    \draw (8pt,-20pt) -- (8pt,44pt);
    \draw (-12pt,0pt) -- (64pt,0pt);
    \draw (-12pt,26pt) -- (64pt,26pt);
    \draw (8pt,44pt) arc (180:0:2.5pt);
    \draw (8pt,-20pt) arc (180:360:2.5pt);
    \draw [densely dotted] (13pt,44pt) -- (13pt,8pt);
    \draw [densely dotted] (13pt,-20pt) -- (13pt,-8pt);
    \draw (46pt,-20pt) -- (46pt,44pt);
    \draw (46pt,44pt) arc (180:0:2.5pt);
    \draw (46pt,-20pt) arc (180:360:2.5pt);
    \draw [densely dotted] (51pt,44pt) -- (51pt,8pt);
    \draw [densely dotted] (51pt,-20pt) -- (51pt,-8pt);
    \filldraw[thick,fill=white] (0pt,18pt) rectangle (16pt,34pt);
    \node at (8pt,26pt) {$U$};
    \filldraw[thick,fill=white] (0pt,-8pt) rectangle (16pt,8pt);
    \node at (8pt,0pt) {$M$};
    \filldraw[thick,fill=white] (38pt,18pt) rectangle (54pt,34pt);
    \node at (46pt,26pt) {$U$};
    \filldraw[thick,fill=white] (38pt,-8pt) rectangle (54pt,8pt);
    \node at (46pt,0pt) {$M$};
    \filldraw [red,fill=red] (-9pt,26pt) circle (1pt) node[left] {$\sigma^z$};
    \filldraw [red,fill=red] (61pt,26pt) circle (1pt) node[right] {$\sigma^y$};
    \filldraw [red,fill=red] (26pt,26pt) circle (1pt) node[above] {$\sigma^x$};
    \end{tikzpicture}\\ 
    &=\ii\begin{tikzpicture}[scale=1,baseline={([yshift=-2pt]current bounding box.center)}]
    \draw (8pt,-20pt) -- (8pt,44pt);
    \draw (-12pt,0pt) -- (64pt,0pt);
    \draw (-12pt,26pt) -- (64pt,26pt);
    \draw (8pt,44pt) arc (180:0:2.5pt);
    \draw (8pt,-20pt) arc (180:360:2.5pt);
    \draw [densely dotted] (13pt,44pt) -- (13pt,8pt);
    \draw [densely dotted] (13pt,-20pt) -- (13pt,-8pt);
    \draw (46pt,-20pt) -- (46pt,44pt);
    \draw (46pt,44pt) arc (180:0:2.5pt);
    \draw (46pt,-20pt) arc (180:360:2.5pt);
    \draw [densely dotted] (51pt,44pt) -- (51pt,8pt);
    \draw [densely dotted] (51pt,-20pt) -- (51pt,-8pt);
    \filldraw[thick,fill=white] (0pt,18pt) rectangle (16pt,34pt);
    \node at (8pt,26pt) {$U$};
    \filldraw[thick,fill=white] (0pt,-8pt) rectangle (16pt,8pt);
    \node at (8pt,0pt) {$M$};
    \filldraw[thick,fill=white] (38pt,18pt) rectangle (54pt,34pt);
    \node at (46pt,26pt) {$U$};
    \filldraw[thick,fill=white] (38pt,-8pt) rectangle (54pt,8pt);
    \node at (46pt,0pt) {$M$};
    \filldraw [red,fill=red] (-9pt,26pt) circle (1pt) node[left] {$\sigma^z$};
    \filldraw [red,fill=red] (61pt,26pt) circle (1pt) node[right] {$\sigma^y$};
    \filldraw [red,fill=red] (26pt,26pt) circle (1pt) node[above] {$\sigma^x$};
    \filldraw [red,fill=red] (46pt,13pt) circle (1pt) node[left] {$\sigma^x$};
    \filldraw [red,fill=red] (46pt,-14pt) circle (1pt) node[left] {$\sigma^x$};
    \filldraw [red,fill=red] (-9pt,0pt) circle (1pt) node[left] {$V_X$};
    \filldraw [red,fill=red] (61pt,0pt) circle (1pt) node[right] {$V_X^{-1}$};
    \end{tikzpicture}\\
    &=-\begin{tikzpicture}[scale=1,baseline={([yshift=-2pt]current bounding box.center)}]
    \draw (8pt,-20pt) -- (8pt,44pt);
    \draw (-12pt,0pt) -- (64pt,0pt);
    \draw (-12pt,26pt) -- (64pt,26pt);
    \draw (8pt,44pt) arc (180:0:2.5pt);
    \draw (8pt,-20pt) arc (180:360:2.5pt);
    \draw [densely dotted] (13pt,44pt) -- (13pt,8pt);
    \draw [densely dotted] (13pt,-20pt) -- (13pt,-8pt);
    \draw (46pt,-20pt) -- (46pt,44pt);
    \draw (46pt,44pt) arc (180:0:2.5pt);
    \draw (46pt,-20pt) arc (180:360:2.5pt);
    \draw [densely dotted] (51pt,44pt) -- (51pt,8pt);
    \draw [densely dotted] (51pt,-20pt) -- (51pt,-8pt);
    \filldraw[thick,fill=white] (0pt,18pt) rectangle (16pt,34pt);
    \node at (8pt,26pt) {$U$};
    \filldraw[thick,fill=white] (0pt,-8pt) rectangle (16pt,8pt);
    \node at (8pt,0pt) {$M$};
    \filldraw[thick,fill=white] (38pt,18pt) rectangle (54pt,34pt);
    \node at (46pt,26pt) {$U$};
    \filldraw[thick,fill=white] (38pt,-8pt) rectangle (54pt,8pt);
    \node at (46pt,0pt) {$M$};
    \filldraw [red,fill=red] (-9pt,26pt) circle (1pt) node[left] {$\sigma^z$};
    \filldraw [red,fill=red] (61pt,26pt) circle (1pt) node[right] {$\sigma^z$};
    \filldraw [red,fill=red] (-9pt,0pt) circle (1pt) node[left] {$V_X$};
    \filldraw [red,fill=red] (61pt,0pt) circle (1pt) node[right] {$V_X^{-1}$};
    \end{tikzpicture}\label{TNsteps}
\end{align}
In the first step, we apply the gauge symmetry condition to both $U$ tensors. From the first to the second line, we use the $\Z_2^C$ symmetry of the $M^2$ tensor given in Eq.~\eqref{Rsym-MPO}. For the next step, we apply the $\Z_2^X$ symmetry of the $U$ tensor on the right, and eliminate the remaining $\sigma^x$ on the physical indices.

Therefore all eigenvalues of $M_U^2$ must come in pairs $\pm\lambda$, including the largest one. Thus we have
\begin{align}
 G(\pi)&\approx\fkh{1+(-1)^{N/2}}\lambda_{\rm max}^N\\
 &=\begin{cases}
    2\lambda_{\rm max}^N & N\equiv 0\pmod{4}\\
    0 & N\equiv 2\pmod{4}
\end{cases}.   
\end{align}
From this result we immediately see that $\gamma(\pi)=\ln 2$.

Now if the largest eigenvalues of $M_U^2$ are $\pm \lambda_{\rm max}$, then for $M_U$ they must be $\lambda_{\rm max}'$ and $\pm \ii\lambda_{\rm max}'$, where the $\pm$ sign can not be directly fixed by this argument. However, this sign ambiguity does not affect $\abs{G(\pi)}$, and one finds that 
\begin{equation}
    \gamma(\pi)=\ln 2\abs{\cos \frac{N\pi}{4}}.
\end{equation}
In particular, $\gamma(\pi)=\ln \sqrt{2}$ for odd $N$.

We note that for the $N\equiv 0\pmod{4}$ case, the argument so far only relies on the MPO invariant under $T^2$. With a doubled unit cell, the filling factor of the $\U$ charge is $1/2$ and the value of $\gamma(\pi)$ is the same as that of the $\mathrm{O}(2)$ LSM case.

Let us now show that the same holds assuming only $T^4$, which is beyond the LSM case. Denote by $S$ the transfer operator with four-site unit cells.  Diagrammatically, $S$ can be represented as
\begin{align}
    S=\begin{tikzpicture}[scale=1,baseline={([yshift=-2pt]current bounding box.center)}]
    \draw (-8pt,0pt) -- (108pt,0pt);
    \draw (-8pt,26pt) -- (108pt,26pt);
    \draw (8pt,-20pt) -- (8pt,44pt);
    \draw (8pt,44pt) arc (180:0:2.5pt);
    \draw (8pt,-20pt) arc (180:360:2.5pt);
    \draw [densely dotted] (13pt,-20pt) -- (13pt,44pt);
    \draw (36pt,-20pt) -- (36pt,44pt);
    \draw (36pt,44pt) arc (180:0:2.5pt);
    \draw (36pt,-20pt) arc (180:360:2.5pt);
    \draw [densely dotted] (41pt,-20pt) -- (41pt,44pt);
    \draw (64pt,-20pt) -- (64pt,44pt);
    \draw (64pt,44pt) arc (180:0:2.5pt);
    \draw (64pt,-20pt) arc (180:360:2.5pt);
    \draw [densely dotted] (69pt,-20pt) -- (69pt,44pt);
    \draw (92pt,-20pt) -- (92pt,44pt);
    \draw (92pt,44pt) arc (180:0:2.5pt);
    \draw (92pt,-20pt) arc (180:360:2.5pt);
    \draw [densely dotted] (97pt,-20pt) -- (97pt,44pt);
    \filldraw[thick,fill=white] (0pt,18pt) rectangle (16pt,34pt);
    \node at (8pt,26pt) {$U$};
    \filldraw[thick,fill=white] (28pt,18pt) rectangle (44pt,34pt);
    \node at (36pt,26pt) {$U$};
    \filldraw[thick,fill=white] (56pt,18pt) rectangle (72pt,34pt);
    \node at (64pt,26pt) {$U$};
    \filldraw[thick,fill=white] (84pt,18pt) rectangle (100pt,34pt);
    \node at (92pt,26pt) {$U$};
    \filldraw[thick,fill=white] (0pt,-8pt) rectangle (100pt,8pt);
    \node at (50pt,0pt) {$M'$};
    \end{tikzpicture}.
\end{align}
Here $M'$ is the tensor of the MPO representation of $\rho$ with four sites in a unit cell. In the fully translation invariant case, we have $S=M_U^4$. Since $\rho$ is invariant under $X_{\rm even}$ ($X_{\rm odd}$), the action $X_{\rm even}$ ($X_{\rm odd}$) can be pushed to the virtual space, which will be denoted by $V_{\rm even}$ ($V_{\rm odd}$). Importantly, $V_{\rm even}$ and $V_{\rm odd}$ must commute, otherwise the transfer matrix obtained by contracting the physical indices of $M'$ would have degenerate spectrum, contradicting the short-range nature of $\rho$.

Following steps very similar to those in Eq.~\eqref{TNsteps}, one can prove
\begin{equation}
\begin{split}
    (\sigma^z\otimes V_{\rm even}) S = S(\sigma^z\otimes V_{\rm even}),\\
    (\sigma^y\otimes V_{\rm odd}) S = S(\sigma^y\otimes V_{\rm odd}).
\end{split}
\end{equation}
Basically, $S$ is invariant under the virtual $\Z_2\times\Z_2$ symmetry generated by $Z\otimes V_{\rm even}$ and $Y\otimes V_{\rm odd}$. Because $V_{\rm even}$ and $V_{\rm odd}$ commute, the virtual states of $S$ form a projective representation of $\Z_2\times\Z_2$ symmetry, and thus the eigenvalues of $S$ are at least twofold degenerate. Again, generically we expect there is no further degeneracy in the spectrum of $S$, and with this assumption we obtain $\gamma(\pi)=\ln 2$.

Heuristically, the result follows from the fact that the unitary $\euler^{\ii \pi Q}$ is a $\Z_2\times\Z_2$ SPT entangler.

\section{Crossover behavior in finite-size systems}
Thus far, all our results are obtained in the limit of $N\rightarrow \infty$, at a fixed temperature. However, for finite system sizes, there may be crossover behavior when the value of $\gamma(\theta)$ is different from the true infinite-size one. One possible scenario for this to happen is when the Hamiltonian is gapped and the ground state has a well-defined charge under the symmetry, leading to suppressed charge fluctuations at very low temperatures. In this section, we will analyze the crossover phenomenon in a simple example. 


\begin{figure}[ht]
    \centering
    \includegraphics[width=\columnwidth]{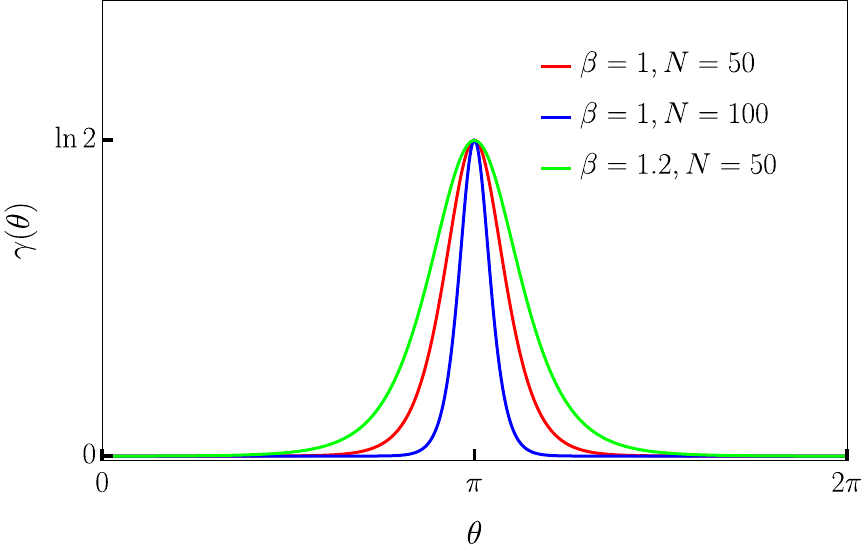}
    \caption{$\gamma(\theta)$ for different temperature $\beta$ and system size $N$. Here, $N=50$ means that we fit $\ln G(\theta)$ using data points near $N=50$.}
    \label{fig_gapped1}
\end{figure}

\begin{figure}[ht]
  \centering
  \includegraphics[width=\columnwidth]{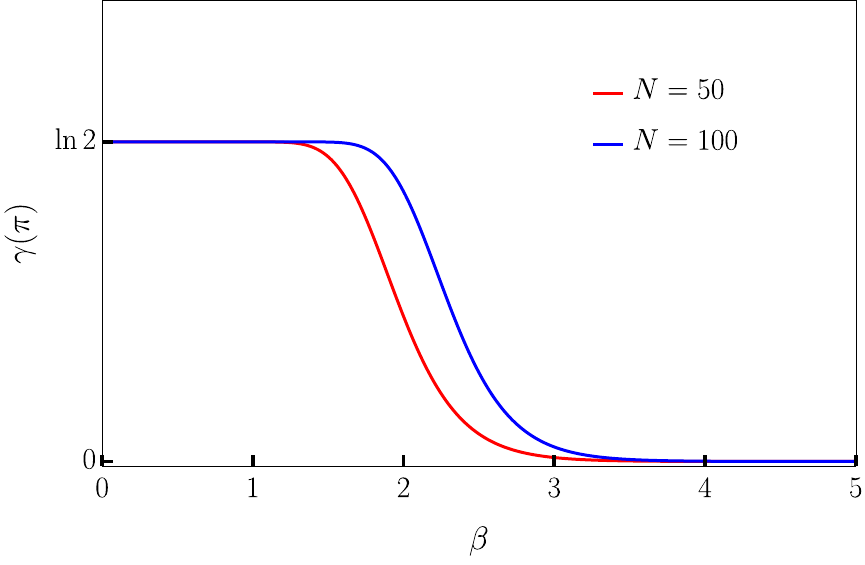}
  \caption{$\gamma(\pi)$ as a function of $\beta$ for different system size $N$.}
  \label{fig_gapped2}
\end{figure}

We will consider the XXZ model described in Eq.~\eqref{XXZ} in the large $\Delta$ limit. For $\Delta>1$, the ground state is in the gapped antiferromagnetic phase. When $\Delta\gg 1$, we can effectively reduce the problem to the antiferromagnetic Ising model:
\begin{align}
    H=\Delta\sum_{n=1}^N \sigma_n^z \sigma_{n+1}^z.
\end{align}
We have
\begin{align}
    G(\theta)=\Tr \euler^{-\beta\sum_n \sigma_n^z \sigma_{n+1}^z} \euler^{\frac12\ii\theta\sum_n \sigma_n^z }.\label{gapped_eq_G}
\end{align}
Here we have absorbed $\Delta$ into the definition of $\beta$.

The evaluation of $G(\theta)$ is essentially the same as the calculation in Sec.~\ref{LG_HighT}. The only difference is that the coupling here is real and we have an imaginary magnetic field.
 We find that the eigenvalues of the transfer matrix in Eq.~\eqref{gapped_eq_G} are given by
\begin{align}
    z_{\pm}(\beta,\theta)=\euler^{-\beta}\kh{\cos\frac{\theta}{2}\pm\sqrt{\euler^{4\beta}-\sin^2\frac{\theta}{2}}}.
\end{align}
Then, $G(\theta)$ is 
\begin{equation}
    G(\theta)=\frac{z_+(\beta,\theta)^N + z_-(\beta,\theta)^N}{z_+(\beta,0)^N + z_-(\beta,0)^N}.
\end{equation}

We can numerically extract $\gamma(\theta)$ by performing a linear fit of $\ln G(\theta)$ with varying system size $N$. In Fig.~\ref{fig_gapped1}, we plot the numerical results of $\gamma(\theta)$ for different temperature and system size. It is evident that $\gamma(\theta)$ exhibits a peak at $\theta=\pi$ with a finite spread in finite systems at finite temperature. When we increase the system size or increase the temperature, the spread gets smaller. In the limit of infinite system size or infinite temperature, $\gamma(\theta)$ will agree with our previous result Eq.~\eqref{gamma-w-XX}.

In Fig.~\ref{fig_gapped2}, we present the numerical results of $\gamma(\pi)$ as a function of temperature for two different system sizes. The plot reveals a cross-over temperature, above which the system exhibits the universal behavior of $\gamma(\pi)$. The cross over temperature decreases as the system size increases.

This crossover can be understood analytically. For even $N$, we find that
\begin{equation}
\begin{split}
    G(\pi)&=\frac{2 (\sqrt{2\sinh 2\beta})^N}{(2\cosh \beta)^N+(2\sinh\beta)^N}\\
    &=\frac{2}{1+\tanh^N\beta} \left(\frac{\sqrt{2\sinh 2\beta}}{2\cosh\beta}\right)^N.
\end{split}
\end{equation}
We may regard the prefactor $\ln\frac{2}{1+\tanh^N\beta}$ as the finite-size value of $\gamma(\pi)$. This provides an explanation for the crossover behavior observed in Fig.~\ref{fig_gapped2}: the crossover occurs when $\tanh^N\beta =1-\epsilon$ for $\epsilon\ll 1$, or $\beta \sim \ln \frac{N}{\epsilon}$.

\section{Conclusions and Discussions}
In this work, we study the constant correction $\gamma(\theta)$ in the thermal expectation value $G(\theta)$ of $\euler^{\ii\theta Q}$, for a (1+1)d periodic lattice system with $\U$ symmetry. We show that $\gamma(\pi)$ becomes quantized in two systems where the $\U$ symmetry has a type-III mixed anomaly with other global symmetry of the system. The value of $\gamma(\pi)$ is closely linked to the symmetry-protected degeneracy of $\U$ symmetry defect. Without such type of anomaly, $\gamma(\pi)$ can depend on microscopic parameters, such as the filling factor. We also provide field-theoretical understandings of these results when the system can be described by a CFT.

An important question left for future work is to more systematically study the relation between the quantization of $\gamma$ and type-III 't Hooft anomaly, beyond the specific example considered in this work. From the CFT perspective, the universal contribution $\gamma$ for a given symmetry operator $g$ is given by $\ln d_g$, where $d_g$ is the degeneracy of the $g$ defect. If $g$ has a type-III mixed anomaly, it means that the $g$ defect transforms projectively under the remaining symmetry $Z_g$ (the centralizer of $g$). The projective class is determined by the anomaly 3-cocycle, as explained in Appendix~\ref{type-anomaly}. We conjecture that generically $d_g$ is the minimal dimension of the irreducible representation in the same projective class.

Following this line of thought, one expects that similar results should hold for fermionic systems. For example, in a (1+1)d fermionic system with $\Z_2\times\Z_2^F$ symmetry ($\Z_2^F$ stands for the fermion parity conservation), the 't Hooft anomaly is classified by $\Z_8$. The generator of the $\Z_8$ is characterized by the $\Z_2$ symmetry defect carrying a Majorana zero mode. Thus we expect that if one measures the expectation value of the total $\Z_2$ charge in a thermal state, $\gamma$ should take a universal value $\gamma=\ln \sqrt{2}$. Another system with a mathematically similar anomaly is a translation-invariant chain of Majorana modes~\cite{FermionLSM}, where the lattice translation has a mixed anomaly with fermion parity. In this case, we expect that the thermal expectation value of the translation operator contains a universal correction $\ln \sqrt{2}$.

An obvious direction for future works is to generalize the results to higher dimensions. It is not difficult to see that the MPO argument for $\gamma(\pi)=\ln 2$ can be generalized to two dimensions, assuming a PEPO representation of the thermal density matrix. It will be worth investigating other classes of systems, such as Fermi liquid or quantum critical points, or systems with other types of 't Hooft anomalies.

\emph{Note added:} During the finalization of the manuscript, we became aware of closely related works \cite{Zang} and \cite{WangFCS}, which study the universal feature in $\alpha(\theta)$. In particular, \cite{Zang} showed that the $\mathrm{O}(2)$ LSM anomaly leads to the cusp at $\theta=\pi$ in $\alpha(\theta)$.

\section{Acknowledgements}
We thank Yingfei Gu, Shenghan Jiang and Pengfei Zhang for helpful conversations and communications on their unpublished works. M.C. is particularly grateful for Hong-Hao Tu for collaboration on a previous related project and many enlightening discussions on tensor networks. M.C. would like to acknowledge NSF for support under award number DMR-1846109.

\bibliography{tdp}

\onecolumngrid

\appendix

\section{\texorpdfstring{$G(\theta)$ of the $c=1$ free boson CFT}{G(θ) of the c=1 free boson CFT}}\label{cft}
Suppose the total charge $Q=t_m Q_m+t_w Q_w$, then
\begin{equation}
    G(\theta)=\Tr \euler^{\ii\theta Q}\euler^{-\beta H_0(\eta_m,\eta_w)} = G_m(\theta_m)G_w(\theta_w).
\end{equation}
Here we have defined $\theta_m=t_m\theta$ and $\theta_w=t_w\theta$. We have factorized the sum over $Q_m$ and $Q_w$, so in the following we only consider $G_m(\theta_m)$.  More explicitly:
\begin{equation}
    G_m(\theta_m)=\frac{\sum_{Q_m\in \mathbb{Z}} \euler^{\ii\theta_m Q_m}\euler^{-\frac{\beta}{2R^2}(Q_m+\eta_w)^2}} {\sum_{Q_m\in \mathbb{Z}} \euler^{-\frac{\beta}{2R^2}(Q_m+\eta_w)^2}} = \frac{\vartheta \kh{\frac{\theta_m}{2\pi}+\frac{\ii\beta\eta_w}{2\pi R^2}, \frac{\ii\beta}{2\pi R^2}}}{\vartheta \kh{\frac{\ii\beta\eta_w}{2\pi R^2}, \frac{\ii\beta}{2\pi R^2}}},
\end{equation}
where the theta function is defined as
\begin{equation}
    \vartheta(z,\tau)=\sum_{n\in \mathbb{Z}} \euler^{\ii\pi \tau n^2+2\pi \ii z n}.
\end{equation}
Using the S transformation
\begin{equation}
    \vartheta(z,\tau)=\frac{1}{\sqrt{-\ii\tau}}\euler^{-\ii\pi \frac{z^2}{\tau}}\vartheta\kh{\frac{z}{\tau},-\frac{1}{\tau}},
\end{equation}
we find
\begin{align}
    \frac{\vartheta \kh{\frac{\theta_m}{2\pi}+\frac{\ii\beta\eta_w}{2\pi R^2}, \frac{\ii\beta}{2\pi R^2}}}{\vartheta \kh{\frac{\ii\beta\eta_w}{2\pi R^2}, \frac{\ii\beta}{2\pi R^2}}} &= \euler^{-\ii\theta_m \eta_w}\euler^{-\frac{R^2\theta_m^2}{2\beta}}
    \frac{\sum_n \exp\kh{-\frac{2\pi^2 R^2}{\beta}(n^2-\frac{\theta_m}{\pi}n)+2\pi \ii \eta_w n}} {\sum_n \exp\kh{-\frac{2\pi^2 R^2}{\beta}n^2+2\pi \ii \eta_w n}}\\
    &\approx  \euler^{-\ii\theta_m \eta_w}\euler^{-\frac{R^2\theta_m^2}{2\beta}} {\sum_n \exp\kh{-\frac{2\pi^2 R^2}{\beta}(n^2-\frac{\theta_m}{\pi}n)+2\pi \ii \eta_w n}} 
\end{align}
In the second line, we consider the $\beta\rightarrow 0$ limit and keep only the leading terms in the sum.

For $0\leq \theta_m<\pi$, we have 
\begin{equation}
    \sum_n \exp\kh{-\frac{2\pi^2 R^2}{\beta}(n^2-\frac{\theta_m}{\pi}n)+2\pi \ii \eta_w n} \approx 1,
\end{equation}
for $\pi<\theta_m<2\pi$:
\begin{equation}
    \sum_n \exp\kh{-\frac{2\pi^2 R^2}{\beta}(n^2-\frac{\theta_m}{\pi}n)+2\pi \ii \eta_w n} \approx \euler^{2\pi \ii\eta_w} \exp\kh{\frac{2\pi R^2(\theta_m-\pi)}{\tilde{\beta}}},
\end{equation}
and for $\theta_m=\pi$:
\begin{equation}
    \sum_n \exp\kh{-\frac{2\pi^2 R^2}{\beta}(n^2-n)+2\pi \ii \eta_w n} \approx  1+\euler^{2\pi \ii\eta_w}=2\euler^{\ii\pi \eta_w}\cos \pi \eta_w.
\end{equation}

Therefore we find 
\begin{equation}
\begin{split}
    G_m(\theta_m)\approx \euler^{-\ii[\theta_m]_\pi\eta_w} \euler^{-\frac{1}{2\beta} R^2[\theta_m]^2},\quad\theta_m\neq \pi\\
    G_m(\pi)\approx 2\cos (\eta_w\pi) \euler^{-\frac{1}{2\beta} \pi^2 R^2},\quad\theta_m=\pi.
\end{split}
\end{equation}

\section{\texorpdfstring{$G(\theta)$}{G(θ)} for XX spin chain}\label{app:XX}
In this section, we provide a detailed derivation of $G(\theta)$ for the XX spin chain. First, we define several quantities that frequently appear in the calculations below. $Z_{ss^{\prime}}(\theta,\beta)$ is defined as
\begin{equation}
    Z_{ss^{\prime}}(\theta,\beta) = \prod_{j=\frac{s^{\prime}}{2}}^{N-1+\frac{s^{\prime}}{2}}\kh{1+(-1)^s\euler^{\ii\theta}\euler^{4\beta \cos\frac{2\pi j}{N}}}.
\end{equation}
Essentially, $Z_{ss^{\prime}}$ is the partition function of free fermions with periodic ($s'=0$) or antiperiodic ($s'=1$) boundary conditions, projected to the fermion parity $(-1)^s$ sector.

We will approximate the discrete products in $Z_{ss^{\prime}}$ by integrals. Define
\begin{align}
    I(\theta,\beta)&=\int_0^{2\pi}\ln\kh{1+\euler^{\ii{\theta}}\euler^{4\beta \cos x}}\diff x\label{Idef1}.
\end{align}
A crucial property of $I(\theta,\beta)$ is that, for $-\pi<\theta\leq\pi$,
\begin{equation}
    \Im I(\theta,\beta)=\pi\theta.
\end{equation}
We show in Appendix~\ref{app:EM} that for $\theta\neq\pi$, we have
\begin{align}
    \sum_{j=\frac{s^{\prime}}{2}}^{N-1+\frac{s^{\prime}}{2}} \ln \left(1+\euler^{\ii\theta}\euler^{4\beta \cos \frac{2\pi j}{N}}\right)  =\frac{N}{2\pi}\int_{0}^{2\pi}\ln (1+\euler^{\ii\theta}\euler^{4\beta \cos x})\diff x + \bO(N^{-1}).
    \label{logsum1}
\end{align}
As a result,
\begin{equation}
    Z_{0s^{\prime}}(\theta,\beta)\approx  \exp\fkh{\frac{N}{2\pi}I(\theta,\beta)} = \euler^{\frac12 \ii [\theta]_\pi N}\exp\fkh{\frac{N}{2\pi}\Re I(\theta,\beta)}.
    \label{Z0s}
\end{equation}
For $\theta=\pi$ and $N\equiv 0\pmod{4}$, we instead have
\begin{equation}
   \sum_{j=\frac12}^{\frac{N}{4}-\frac12} \ln (\euler^{4\beta \cos \frac{2\pi j}{N}}-1)=\frac{N}{2\pi}\int_{0}^{\frac{\pi}{2}}\ln (\euler^{4\beta \cos x}-1)\diff x+\ln \sqrt{2} + \bO(N^{-1}).
   \label{logsum2}
\end{equation}

We also need the following inequality between the integrals. Note that
\begin{align}
    \abs{1+\euler^{\ii\theta}\euler^{4\beta \cos x}}^2=1+\euler^{8\beta \cos x}+2\cos \theta\euler^{4\beta\cos x}\leq (1+\euler^{4\beta\cos x})^2,
\end{align}
which implies $I(0,\beta)>\Re I\kh{\theta,\beta}$ for $0<\theta<2\pi$.

\subsection{Half filling}
Using Jordan-Wigner transformation, the periodic XX Hamiltonian can be mapped to a free fermion Hamiltonian with two sectors. One has odd number of fermions and periodic boundary condition for the fermions, and the other one has even number of fermions and antiperiodic boundary condition for the fermions. We can write $\Tr\euler^{\ii\theta Q}\euler^{-\beta H}$ as
\begin{equation}
    \Tr\euler^{\ii\theta Q}\euler^{-\beta H}=\frac12\fkh{Z_{00}(\theta,\beta)-Z_{10}(\theta,\beta)+Z_{01}(\theta,\beta)+Z_{11}(\theta,\beta)}.
\end{equation}

As shown in Eq.~\eqref{Z0s}, for $0<\theta<\pi$, we have
\begin{equation}
    \Tr\euler^{\ii\theta Q}\euler^{-\beta H}\approx\exp\fkh{\frac{N}{2\pi}I\kh{\theta,\beta}}
    =\exp\kh{\frac12 \ii\theta N}\exp\fkh{\frac{N}{2\pi}\Re I\kh{\theta,\beta}}.
\end{equation}
For $\theta=\pi$,
\begin{equation}
    \Tr\euler^{\ii\pi Q}\euler^{-\beta H}=\frac12\fkh{-Z_{00}(0,\beta)+Z_{10}(0,\beta)+Z_{01}(0,\beta)+Z_{11}(0,\beta)}.\label{eq_xx_ipiQ}
\end{equation}
We have proved that $\Tr\euler^{\ii\pi Q}\euler^{-\beta H}=0$ for $N$ odd, so we only need to consider the $N$ even case. $Z_{00}(0,\beta)$ and $Z_{01}(0,\beta)$ can be approximated by Eq.~\eqref{Z0s}, and they cancel each other. The other two terms need to be considered separately for $N\equiv 0 \pmod{4}$ and $N\equiv 2 \pmod{4}$. For $N\equiv 0\pmod{4}$, we have
\begin{align}
    Z_{10}(0,\beta)=\prod_{j=0}^{N-1} (1-\euler^{4\beta \cos\frac{2\pi j}{N}})=0,
\end{align}
and by Eq.~\eqref{logsum2}
\begin{align}
    Z_{11}(0,\beta)&=\prod_{j=\frac{1}{2}}^{N-\frac{1}{2}} (1-\euler^{4\beta \cos\frac{2\pi j}{N}})=\left[\prod_{j=\frac{1}{2}}^{\frac{N}{4}-\frac{1}{2}} (\euler^{4\beta \cos\frac{2\pi j}{N}}-1)\right]^2 \left[\prod_{j=\frac{N}{4}+\frac12}^{\frac{N}{2}-\frac{1}{2}} (1-\euler^{4\beta \cos\frac{2\pi j}{N}})\right]^2\\
    &\approx 4\exp\fkh{\frac{N}{2\pi}\Re I(\pi,\beta)}.
\end{align}
When $N\equiv 2\pmod{4}$, we have the opposite: $Z_{11}(0,\beta)=0$ and $Z_{10}\approx -4\exp\fkh{\frac{N}{2\pi}\Re I(\pi,\beta)}$. Thus $\Tr\euler^{\ii\pi Q}\euler^{-\beta H}\approx 2(-1)^{\frac{N}{2}}\exp\fkh{\frac{N}{2\pi}\Re I(\pi,\beta)}$. Note the prefactor $2$, which arises from $\bO\kh{N^{-1}}$ correction to approximating the sum by an integral, is key to the topological correction $\gamma(\pi)$ in XX spin chain.

The partition function $\Tr\euler^{-\beta H}$ contains the same terms that appear in $\Tr\euler^{\ii\pi Q}\euler^{-\beta H}$ with different signs.

In conclusion, for $0<\theta<\pi$, we have
\begin{align}
    G(\theta)\approx \euler^{\frac12\ii\theta N} \euler^{-\frac{N}{2\pi}\fkh{I(0,\beta)-\Re I(\theta,\beta)}}.
\end{align}
While for $\theta=\pi$,
\begin{align}
    G(\pi)&\approx 2\cos\kh{\frac{N\pi}{2}}\euler^{-\frac{N}{2\pi}\fkh{I(0,\beta)-\Re I(\pi,\beta)}}.
\end{align}

\subsection{Away from half filling}
With a nonzero chemical potential $\mu$, the partition function of XX spin chain can be written as
\begin{align}
    \Tr\euler^{\ii\theta Q}\euler^{-\beta H}=&
        \frac{1}{2}\fkh{\prod_{j=0}^{N-1}\kh{1+\euler^{4\beta\cos \frac{2\pi j}{N}-4\beta\cos(\pi\nu)+\ii\theta}}
        -\prod_{j=0}^{N-1}\kh{1-\euler^{4\beta\cos \frac{2\pi j}{N}-4\beta\cos(\pi\nu)+\ii\theta}}}\notag\\
        &+\frac{1}{2}\fkh{\prod_{j=\frac{1}{2}}^{N-\frac{1}{2}}\kh{1+\euler^{4\beta\cos \frac{2\pi j}{N}-4\beta\cos(\pi\nu)+\ii\theta}}
        +\prod_{j=\frac{1}{2}}^{N-\frac{1}{2}}\kh{1-\euler^{4\beta\cos \frac{2\pi j}{N}-4\beta\cos(\pi\nu)+\ii\theta}}},
\end{align}
where $\nu=\frac{1}{\pi}\arccos\kh{-\frac{\mu}{4}}$ is the filling fraction at zero temperature.

We can approximate the products by integrals as
\begin{align}
    \Tr\euler^{-\beta H}
        \approx&\exp\fkh{\frac{N}{2\pi}I(0,\beta,\nu)},
\end{align}
where
\begin{align}
    I(\theta,\beta,\nu)&=\int_0^{2\pi}\ln\kh{1+\euler^{\ii\theta}\euler^{4\beta\cos x-4\beta\cos(\pi\nu)}} \diff x.
\end{align}
For $0<\theta<\pi$, we have
\begin{align}
    \abs{\Tr\euler^{\ii\theta Q}\euler^{-\beta H}}\approx \exp\fkh{\frac{N}{2\pi}\Re I(\theta,\beta,\nu)}.
\end{align}
The phase factor now depends on all the parameters $\theta,\beta$ and $\nu$. For $\theta=\pi$, we show in Appendix~\ref{app:EM} that the constant correction relating $I(\pi,\beta,\nu)$ and $\ln\prod\kh{1-\euler^{4\beta\cos\frac{2\pi j}{N}-4\beta\cos(\pi\nu)}}$ is now $2\ln\fkh{2\sin(\pi \delta)}$, where $\delta$ (or $1-\delta$) is the minimal distance between $j$ and the singular point $x=\frac{N\nu}{2}$. For $j$ taking integer values, we can take $\delta=\fkh{\frac{N\nu}{2}}$, which is the decimal part of $\frac{N\nu}{2}$. For $j$ taking half-integer values, we can take $\delta=\abs{\fkh{\frac{N\nu}{2}}-\frac{1}{2}}$.
Thus
\begin{align}
    \Tr\euler^{\ii\pi Q}\euler^{-\beta H}&\approx -2\sin^2\kh{\fkh{\frac{N\nu}{2}}\pi}\exp\fkh{\Re I(\pi,\beta,\nu)}+2\sin^2\kh{\fkh{\frac{N\nu}{2}}\pi-\frac{1}{2}\pi}\exp\fkh{\Re I(\pi,\beta,\nu)}\\
    &=2\cos\kh{\nu N\pi}\exp\fkh{\frac{N}{2\pi}\Re I(\pi,\beta,\nu)}.
\end{align}

In conclusion, for $0<\theta<\pi$, we have
\begin{equation}
    \abs{G(\theta)}\approx\euler^{-\frac{N}{2\pi}\fkh{I(0,\beta,\nu)-\Re I(\theta,\beta,\nu)}}.
\end{equation}
While for $\theta=\pi$,
\begin{equation}
    G(\pi)\approx 2\cos(\nu N\pi)\euler^{-\frac{N}{2\pi}\fkh{I(0,\beta,\nu)-\Re I(\pi,\beta,\nu)}}.
\end{equation}

\section{\texorpdfstring{$G(\theta)$}{G(θ)} of the Levin-Gu model}\label{app:LG}
\subsection{\texorpdfstring{$N$ even}{N even}}
For $N$ even, the Levin-Gu model can be mapped to a free fermion system with even number of fermions and periodic or antiperiodic boundary condition, we have
\begin{align}
    \Tr\euler^{\ii\theta Q}\euler^{-\beta H}=&
        \frac{\euler^{\frac12\ii\theta N}}{2}\fkh{Z_{00}(-\theta/2,\beta)+Z_{10}(-\theta/2,\beta)+Z_{01}(-\theta/2,\beta)+Z_{11}(-\theta/2,\beta)}.
\end{align}

First, we consider $0<\theta<\pi$. As shown in Eq.~\eqref{Z0s}, 
\begin{align}
    \Tr\euler^{\ii\theta Q}\euler^{-\beta H}&\approx\exp\fkh{\frac{N}{2\pi}I\kh{-\frac{\theta}{2},\beta}+\frac{1}{2}\ii\theta N} +\exp\fkh{\frac{N}{2\pi}I\kh{-\frac{\theta}{2}+\pi,\beta}+\frac{1}{2}\ii\theta N}  \\
    &\approx \exp\fkh{\frac{N}{2\pi}I\kh{-\frac{\theta}{2},\beta}+\frac{1}{2}\ii\theta N}\\
    &= \exp\kh{\frac{1}{4}\ii\theta N}\exp\fkh{\frac{N}{2\pi}\Re I\kh{\frac{\theta}{2},\beta}}.
\end{align}
The second line follows from $\abs{1+\euler^{-\ii\frac{\theta}{2}}\euler^{4\beta \cos x}}>\abs{1-\euler^{-\ii\frac{\theta}{2}}\euler^{4\beta \cos x}}$ for any $0<\theta<\pi$. From this result we immediately see $\omega(\theta)=\exp\kh{\frac{1}{4}\ii\theta N}$ for $0<\theta<\pi$.

For $\theta=\pi$, from Eq.~\eqref{Z0s}, we have
\begin{align}
    Z_{0s^{\prime}}(\pm\pi/2,\beta)=Z_{1s^{\prime}}(\mp\pi/2,\beta)=\exp\fkh{\frac{N}{2\pi}I\kh{\pm\frac{\pi}{2},\beta}}.
\end{align}
Crucially, the integrals satisfy $\Re I(\frac{\pi}{2},\beta)=\Re I(-\frac{\pi}{2},\beta)$.
Therefore we have 
\begin{equation}
    \Tr\euler^{\ii\pi Q}\euler^{-\beta H}\approx 2\cos \kh{\frac{\pi N}{4}} \exp\fkh{\frac{N}{2\pi}\Re I(\frac{\pi}{2},\beta)}.
\end{equation}

The partition function contains the same terms in Eq.~\eqref{eq_xx_ipiQ} discussed in the previous section:
\begin{align}
    \Tr\euler^{-\beta H}=&
        \frac12 \fkh{Z_{00}(0,\beta)+Z_{10}(0,\beta)+Z_{01}(0,\beta)+Z_{11}(0,\beta)}.
\end{align}

Putting everything together, we have found that for $0<\theta<\pi$,
\begin{equation}
    G(\theta)\approx 
    \frac{\exp\kh{\frac{1}{4}\ii\theta N}\exp\fkh{\frac{N}{2\pi}\Re I\kh{\frac{\theta}{2},\beta}}}{\exp\fkh{\frac{N}{2\pi}I(0,\beta)}+2(-1)^{\frac{N}{2}}\exp\fkh{\frac{N}{2\pi}\Re I(\pi,\beta)}}\approx \euler^{\frac{1}{4}\ii\theta N}\euler^{-\frac{N}{2\pi}\fkh{I(0,\beta)-\Re I\kh{\frac{\theta}{2},\beta}}},
\end{equation}
and for $\theta=\pi$,
\begin{equation}
    G(\pi)\approx\frac{2\cos\kh{\frac{\pi N}{4}}\exp\fkh{\frac{N}{2\pi}\Re I(\frac{\pi}{2},\beta)}}{\exp\fkh{\frac{N}{2\pi}I(0,\beta)}+2(-1)^{\frac{N}{2}}\exp\fkh{\frac{N}{2\pi}\Re I(\pi,\beta)}}\approx 2\cos\kh{\frac{\pi N}{4}}\euler^{-\frac{N}{2\pi}\fkh{I(0,\beta)-\Re I(\frac{\pi}{2},\beta)}}.
\end{equation}

\subsection{\texorpdfstring{$N$ odd}{N odd}}
When $N$ is odd, the Levin-Gu model corresponds to a free fermion system with odd number of fermions, and we can write
\begin{align}
    \Tr\euler^{\ii\theta Q}\euler^{-\beta H}
    =\frac{\euler^{\frac{1}{2}\ii\theta N}}{2}\fkh{Z_{00}(-\theta/2,\beta)-Z_{10}(-\theta/2,\beta)+Z_{01}(-\theta/2,\beta)-Z_{11}(-\theta/2,\beta)}.
\end{align}
For $0<\theta<\pi$, we have
\begin{align}
    \Tr\euler^{\ii\theta Q}\euler^{-\beta H}
    &\approx\exp\fkh{\frac{N}{2\pi}I\kh{-\frac{\theta}{2},\beta}+\frac{1}{2}\ii\theta N} -\exp\fkh{\frac{N}{2\pi}I\kh{-\frac{\theta}{2}+\pi,\beta}+\frac{1}{2}\ii\theta N}  \\
    &\approx \exp\fkh{\frac{N}{2\pi}I\kh{-\frac{\theta}{2},\beta}+\frac{1}{2}\ii\theta N}\\
    &=\exp\kh{\frac{1}{4}\ii\theta N}\exp\fkh{\frac{N}{2\pi}\Re I\kh{\frac{\theta}{2},\beta}},
\end{align}
while for $\theta=\pi$, we have
\begin{align}
    \Tr\euler^{\ii\pi Q}\euler^{-\beta H}
    \approx &\fkh{\exp\kh{\frac{\ii\pi N}{4}}-\exp\kh{\frac{3\ii\pi N}{4}}}\exp\fkh{\frac{N}{2\pi}\Re I(\frac{\pi}{2},\beta)}\\
    =& 2(-1)^{\frac{N-1}{2}} \sin \kh{\frac{\pi N}{4}} \exp\fkh{\frac{N}{2\pi }\Re I(\frac{\pi}{2},\beta)}.
\end{align}

The large $N$ behavior of the partition function is still governed by $I(0,\beta)$ as $I(0,\beta)>\Re I(\pi,\beta)$. Moreover, the two terms $Z_{10}(0,\beta)$ and $Z_{11}(0,\beta)$ have opposite sign for $N$ odd and they cancel each other.
Write $N=4m+r$ where $r=1,3$, then $\sin \frac{\pi N}{4}=\sin (m\pi + \frac{r\pi}{4})=(-1)^m \sin \frac{r\pi}{4}=\frac{1}{\sqrt{2}}(-1)^m$.
We conclude that for $0<\theta<\pi$,
\begin{equation}
    G(\theta)\approx
    \frac{\exp\kh{\frac{1}{4}\ii\theta N}\exp\fkh{\frac{N}{2\pi}\Re I\kh{\frac{\theta}{2},\beta}}}{\exp\fkh{\frac{N}{2\pi}I(0,\beta)}}=\euler^{\frac{1}{4}\ii\theta N}\euler^{-\frac{N}{2\pi}\fkh{I(0,\beta)-\Re I\kh{\frac{\theta}{2},\beta}}},
\end{equation}
and for $\theta=\pi$,
\begin{equation}
    G(\pi)\approx\frac{(-1)^{m+\frac{r-1}{2}}\sqrt{2}\exp\fkh{\frac{N}{2\pi}\Re I(\frac{\pi}{2},\beta)}}{\exp\fkh{\frac{N}{2\pi}I(0,\beta)}}
    =(-1)^{m+\frac{r-1}{2}}\sqrt{2} \euler^{-\frac{N}{2\pi}\fkh{I(0,\beta)-\Re I(\frac{\pi}{2},\beta)}}.
\end{equation}

Notice that $\cos \frac{\pi N}{4}=\cos (m\pi+\frac{\pi r}{4})=(-1)^{m}\cos \frac{\pi r}{4}=(-1)^m (-1)^{\frac{r-1}{2}}\frac{1}{\sqrt{2}}$. So the phase factor $(-1)^{m+\frac{r-1}{2}}$ is equal to $\sgn(\cos \frac{\pi N}{4})$ in the main text.

\section{Evaluation of the product}\label{app:EM}
The Euler-MacLaurin formula for a continuously differentiable function $f(x)$ reads
\begin{equation}
    \sum_{i=a}^b f(i)=\int_a^b f(x)\diff x + \frac{f(a)+f(b)}{2}+ R_m.
    \label{E-M}
\end{equation}
The remainder term $R_m$ is given by
\begin{equation}
    R_m=\sum_{k=1}^m \frac{B_{2k}}{(2k)!}[f^{(2k-1)}(b)-f^{(2k-1)}(a)]+\int_a^b P_{2m+1}(x)f^{(2m+1)}(x) \diff x, \: m \in \N.
\end{equation}
Here $B_k$ are the Bernoulli numbers, and the periodized Bernoulli functions $P_k(x)$ is defined as $P_k(x)=B_k(x-[x])$, where $B_k(x)$ is the Bernoulli polynomial.

A particularly useful result is that when $|f'(x)|$ is bounded on $[a,b]$:
\begin{equation}
    \abs{R_0}=\abs{\int_a^b P_{1}(x)f'(x) \diff x}\leq\int_a^b \abs{P_{1}(x)f'(x)}\diff x \leq \int_a^b \abs{f'(x)}\diff x\leq (b-a)\max_{x\in [a,b]} |f'(x)|.
\end{equation}
Here we have used $\abs{P_1(x)}=\abs{x-[x]-\frac12}\leq 1$.

We now use the Euler-MacLaurin formula to study the sum
\begin{equation}
\sum_{j=0}^{N-1} \ln \left(\euler^{\ii\theta}\euler^{\alpha \cos \frac{2\pi j}{N}}+1\right).
\end{equation}
Applying Eq.~\eqref{E-M} with $f(x)=\ln \left(\euler^{\ii\theta}\euler^{\alpha \cos \frac{2\pi x}{N}}+1\right)$, we obtain
\begin{align}
     \sum_{j=0}^{N-1} \ln \left(\euler^{\ii\theta}\euler^{\alpha \cos \frac{2\pi j}{N}}+1\right) &= \int_{0}^{N-1} \ln (\euler^{\alpha \cos \frac{2\pi}{N} x}+1)\diff x  +  \frac12 [\ln (\euler^{\ii\theta}\euler^{{\alpha}}+1) +\ln (\euler^{\ii\theta}\euler^{{\alpha \cos\frac{2\pi}{N}}}+1)]  +R\\
     &\approx\frac{N}{2\pi}\int_{0}^{2\pi-\frac{2\pi}{N}} \ln (\euler^{\ii\theta}\euler^{\alpha \cos x}+1)\diff x  +  \ln (\euler^{\ii\theta}\euler^{{\alpha }}+1) +\bO(N^{-2})+R\\
     &=\frac{N}{2\pi}\left(\int_{0}^{2\pi}   -\int_{2\pi-\frac{2\pi}{N}}^{2\pi}\right)\ln (\euler^{\ii\theta}\euler^{\alpha \cos x}+1)\diff x +   \ln (\euler^{\ii\theta}\euler^{{\alpha }}+1)   + R+\bO(N^{-2}) \\
     &=\frac{N}{2\pi}\int_{0}^{2\pi}\ln (\euler^{\ii\theta}\euler^{\alpha \cos x}+1)\diff x + R + \bO(N^{-2}).  
\end{align}
Now we estimate the remainder term $R$. We observe that (with $y=\frac{2\pi}{N}x$)
\begin{equation}
    \abs{f'(x)}=\frac{2\pi \alpha}{N}\abs{\frac{\euler^{\alpha \cos y}}{\euler^{\ii\theta}\euler^{\alpha \cos y}+1}\sin y}\leq\frac{2\pi \alpha}{N}\frac{\euler^{\alpha \cos y}}{\abs{\euler^{\alpha \cos y}+\euler^{-\ii\theta}}}.
\end{equation}
When the imaginary part of $\euler^{\ii\theta}$ is nonzero, i.e., $\sin\theta\neq 0$, then the denominator is lower-bounded by $\abs{\sin\theta}$ and the numerator is upper-bounded by $\euler^\alpha$, which imply 
\begin{equation}
    |R_0|=\bO(N^{-1}). \label{bound-R}
\end{equation}

Let us now consider $\sin\theta=0$, i.e., $\theta=0$ or $\pi$. For $\theta=0$, we have 
\begin{equation}
\begin{split}
    \abs{f'(x)} &\leq\frac{2\pi \alpha}{N}\frac{\euler^{\alpha \cos y}}{{\euler^{\alpha \cos y}+1}}\leq \frac{2\pi \alpha}{N}.
\end{split}
\end{equation}
Therefore Eq.~\eqref{bound-R} still holds. Thus, for any $\theta\neq \pi\pmod{2\pi}$, we have shown that
\begin{align}
     \sum_{j=0}^{N-1} \ln \left(\euler^{\ii\theta}\euler^{\alpha \cos \frac{2\pi j}{N}}+1\right)  =\frac{N}{2\pi}\int_{0}^{2\pi}\ln (\euler^{\ii\theta}\euler^{\alpha \cos x}+1)\diff x + \bO(N^{-1}).  
\end{align}
It is straightforward to extend the proof to the case where the sum is over half-integer values.

Next we deal with the special $\theta=\pi$ case, where the derivative $f'(x)$ diverges at $x=\pi/2$. In fact, the same is true for any $f^{(2m+1)}(x)$ at $x=\pi/2$, so the formula is not particularly useful for any finite $m$. Thus we formally extend the formula to $m=\infty$:
\begin{equation}
    R_\infty\equiv R_{m=\infty}=\sum_{k=1}^\infty \frac{B_{2k}}{(2k)!}[f^{(2k-1)}(b)-f^{(2k-1)}(a)].
\end{equation}
It should be understood as an asymptotic series expansion of the remainder term. To get a finite result, we will need to resum the series.

First let us assume $N\equiv 0\pmod{4}$. We will consider the sum
\begin{equation}
\sum_{j=\frac12}^{\frac{N}{4}-\frac12} \ln (\euler^{\alpha \cos \frac{2\pi j}{N}}-1).
\end{equation}
Let $f(x)=\ln (\euler^{\alpha \cos \frac{2\pi}{N} x}-1)$, then we have
\begin{align}
    \sum_{j=\frac12}^{\frac{N}{4}-\frac12} \ln (\euler^{\alpha \cos \frac{2\pi j}{N}}-1) &= \int_{\frac12}^{\frac{N}{4}-\frac12} \ln (\euler^{\alpha \cos \frac{2\pi}{N} x}-1)\diff x  +  \frac12 [\ln (\euler^{{\alpha \cos \frac{\pi}{N}}}-1) +\ln (\euler^{{\alpha \sin\frac{\pi}{N}}}-1)]  +R_\infty\\
    &=\frac{N}{2\pi}\int_{\frac{\pi}{N}}^{\frac{\pi}{2}-\frac{\pi}{N}} \ln (\euler^{\alpha \cos x}-1)\diff x  +  \frac12 \fkh{\ln (\euler^{{\alpha }}-1) +\ln \frac{\alpha\pi}{N}}+\bO(N^{-2})+R_\infty\\
    &=\frac{N}{2\pi}\int_{0}^{\frac{\pi}{2}} \ln (\euler^{\alpha \cos x}-1)\diff x+\frac12 + R_\infty + \bO(N^{-2}).  
\end{align}

Now we consider the remainder term. It is not difficult to show that
\begin{equation}
    f^{(2k-1)}\kh{\frac{N}{4}-\frac12}=\left(\frac{2\pi}{N}\right)^{2k-1} \frac{(2k-2)!}{(-\pi/N)^{2k-1}}=-2^{2k-1} (2k-2)!.
\end{equation}
And $f^{(2k-1)}(\frac12)$ is $\bO(N^{-2k-2})$, so it can be ignored.
Therefore
\begin{equation}
    R_\infty=-\sum_{k=1}^\infty\frac{B_{2k}}{2k(2k-1)}2^{2k-1}.
\end{equation}
Here we can see that the correction does not depend on $\alpha$, neither does it depend on the specific form of the spectrum considered. This explains why CFT results agree with results derived from lattice models at any finite temperature.

To resum the asymtotic series, we use the following integral representation of Bernouli numbers:
\begin{equation}
    B_{2k}=4k (-1)^{k+1} \int_0^{\infty} \frac{t^{2k-1}}{\euler^{2\pi t}-1}\diff t.
\end{equation}
Then we find
\begin{equation}
   R_\infty=-2 \int_0^{\infty} \frac{\diff t}{\euler^{2\pi t}-1}\sum_{k=1}^{\infty} (-1)^{k+1}\frac{(2t)^{2k-1}}{2k-1}=- 2\int_0^{\infty} \frac{\arctan 2t}{\euler^{2\pi t}-1}\diff t=-\left(\frac12-\ln\sqrt{2}\right).
\end{equation}
The last integral is given by Binet's second formula for the logarithm of the Gamma function~\cite{whittaker_watson_1996}:
\begin{align}
    2\int_0^{\infty} \frac{1}{\euler^{2\pi t}-1} \arctan \frac{t}{z}\diff t= z-\kh{z-\frac12}\ln z+\ln \Gamma\kh{z}-\ln \sqrt{2\pi}.
\end{align}
Setting $z=\frac12$ gives the desired result.

So putting together we have shown that 
\begin{equation}
   \sum_{j=\frac12}^{\frac{N}{4}-\frac12} \ln (\euler^{\alpha \cos \frac{2\pi j}{N}}-1) =\frac{N}{2\pi}\int_{0}^{\frac{\pi}{2}}\ln (\euler^{\alpha \cos x}-1)\diff x+\ln \sqrt{2} + \bO(N^{-1}).
\end{equation}
Similarly,
\begin{equation}
\begin{split}
     \sum_{j=\frac{N}{4}+\frac12}^{\frac{N}{2}-\frac12} \ln (1-\euler^{\alpha \cos \frac{2\pi j}{N}}) &= 
    \sum_{j=\frac{N}{4}+\frac12}^{\frac{N}{2}-\frac12} \ln \euler^{\alpha \cos \frac{2\pi j}{N}}(\euler^{-\alpha \cos \frac{2\pi j}{N}}-1) \\
    &=\alpha\sum_{j=\frac{N}{4}+\frac12}^{\frac{N}{2}-\frac12} \cos \frac{2\pi j}{N}+\sum_{j=\frac{N}{4}+\frac12}^{\frac{N}{2}-\frac12}\ln (\euler^{-\alpha \cos \frac{2\pi j}{N}}-1)\\
    &=-\frac{\alpha}{2\sin\frac{\pi}{N}}+\sum_{j=\frac12}^{\frac{N}{4}-\frac12}\ln (\euler^{-\alpha \cos \frac{2\pi j}{N}}-1)\\
    &=-\frac{\alpha}{2\sin\frac{\pi}{N}}+ \frac{N}{2\pi}\int_{0}^{\frac{\pi}{2}}\ln (\euler^{\alpha \cos x}-1)\diff x+\ln \sqrt{2} + \bO(N^{-1})\\
    &= \frac{N}{2\pi}\int_{\frac{\pi}{2}}^{\pi} \ln (1-\euler^{\alpha \cos x})\diff x+\ln \sqrt{2} + \bO(N^{-1}).
\end{split}
\end{equation}

A straightforward generalization is to consider $f(x)=\ln\kh{\euler^{\alpha \cos\frac{2\pi}{N}x-\alpha\cos\kh{\pi\nu}}-1}$ and the summation
\begin{equation}
    \sum_{j=\frac12 s^{\prime}}^{N-1+\frac12 s^{\prime}} f(j),
\end{equation}
where $\nu\in(0,1)$ and $s^{\prime}\in\hkh{0,1}$. Now the singular points are at $x_1=\frac{N\nu}{2}$ and $x_2=N-\frac{N\nu}{2}$. Assume the largest $j<x_1$ in the summation is $j_1=x_1-\delta$. Then, we have
\begin{align}
    \sum_{j=\frac12 s^{\prime}}^{x_1-\delta} f(j)&=\int_{\frac12 s^{\prime}}^{x_1-\delta} f(x)\diff x+\frac12 \fkh{f(\frac12s^{\prime})+f(x_1-\delta)}+R_{\infty}+\bO(N^{-2})
    \\&=\int_{0}^{x_1} f(x)\diff x+\kh{\frac12-\delta}\ln\fkh{\alpha \sin(\pi\nu)\frac{2\pi \delta}{N}}+\delta+R_{\infty}+\bO(N^{-2}).
\end{align}
The remainder term $R_{\infty}$ is
\begin{equation}
    R_{\infty}=-\sum_{k=1}^{\infty} \frac{B_{2k}}{2k(2k-1)}\kh{\frac{1}{\delta}}^{2k-1}=-\delta+\kh{\delta-\frac12}\ln \delta+\ln\frac{\sqrt{2\pi}}{\Gamma(\delta)}.
\end{equation}
Thus
\begin{equation}
    \sum_{j=\frac12 s^{\prime}}^{x_1-\delta} f(j)=\int_{0}^{x_1} f(x)\diff x+\kh{\frac12-\delta}\ln\fkh{\alpha \sin(\pi\nu)\frac{2\pi \delta}{N}}+\kh{\delta-\frac12}\ln \delta+\ln\frac{\sqrt{2\pi}}{\Gamma(\delta)}.
\end{equation}
If we add up the constant corrections from four segments, we will find
\begin{equation}
    \sum_{j=\frac12 s^{\prime}}^{N-1+\frac12 s^{\prime}} \ln\abs{\euler^{\alpha \cos\frac{2\pi}{N}j-\alpha\cos\kh{\pi\nu}}-1}=\frac{N}{2\pi}\int_0^{2\pi}\ln\abs{\euler^{\alpha \cos x-\alpha\cos\kh{\pi\nu}}-1}\diff x+2\ln \fkh{2\sin(\pi \delta)}.
\end{equation}

\section{Typology of 't Hooft anomaly in (1+1)d}\label{type-anomaly}
In (1+1)d bosonic/spin systems, 't Hooft anomaly for a global symmetry $G$ is classified by $\H^3(G, \U)$. Namely, each anomaly is uniquely associated with a group cohomology class $[\omega]\in \H^3(G, \U)$.

Anomalies can be partially characterized by symmetry transformation properties of a symmetry defect. More concretely, given $g\in G$, one considers the system with a $g$ defect, which can be viewed as a (0+1)d quantum-mechanical system with $Z_g$ symmetry, where $Z_g=\{h\in G|hg=gh\}$ is the centralizer of $g$. The $Z_g$ symmetry action may be projective, characterized by a 2-cocycle $\omega_g = i_g\omega$ in $\H^2(Z_g, \U)$, where $i_g$ is the slant product. The explicit expression for $i_g\omega$ is given by
\begin{equation}
    (i_g\omega)(h,k)=\frac{\omega(g,h,k)\omega(h,k,g)}{\omega(h,g,k)}.
\end{equation}

If $\omega_g$ is nontrivial for some $g\in G$, the anomaly $\omega$ is said to be type-III. To give an example, consider $G=\Z_2^3$. Label the group elements by $a\equiv (a_1,a_2,a_3)$ where $a_1,a_2,a_3\in\{0,1\}$, and the group multiplication is defined as addition mod $2$. The type-III cocycle is given by 
\begin{equation}
    \omega(a,b,c)=(-1)^{a_1b_2c_3}.
\end{equation}
To see that it is indeed type-III, we compute $i_{(1,0,0)}$:
\begin{equation}
    (i_{(1,0,0)}\omega)(b,c)=(-1)^{b_2c_3}.
\end{equation}
This is the 2-cocycle for the projective representation of the $\Z_2^2$ subgroup generated by $(0,1,0)$ and $(0,0,1)$.

Let us show that the two examples discussed in the main text have type-III anomaly.

We start with the XX spin chain. The Hamiltonian can be modified to have a $\theta=\pi$ defect at the link $N,1$: 
\begin{equation}
\begin{split}\label{SO3defH}
    H(\pi)=-\sum_{n=1}^{N-1} (\sigma_n^x\sigma_{n+1}^x + \sigma_n^y\sigma_{n+1}^y) + \sigma_N^x\sigma_{1}^x + \sigma_N^y\sigma_{1}^y.
\end{split}
\end{equation}
It is clear that $X$ remains a symmetry of $H(\pi)$, but the lattice translation needs to be modified:
\begin{equation}
    T(\pi)=\euler^{\ii \frac{\pi}{2}\sigma^z_1} T=\ii \sigma_1^z T.
\end{equation}
Therefore $T(\pi)X=-XT(\pi)$, implying that the system with a $\Z_2\subset \U$ defect transforms projectively under the $\Z_2^C$ and lattice translation.

Next we consider the Levin-Gu model, introducing a $\theta=\pi$ defect~\cite{Cheng:2022sgb}:
\begin{equation}
  H_{\rm LG}(\pi)=-\sum_{n=1}^{N-1} \mu^z_{n,n+1}(\tau_n^+\tau_{n+1}^- + \text{h.c.})+\mu^z_{N,1}(-\ii\tau_N^+\tau_{1}^- +\text{h.c.}).
  \label{LGHsigma}
\end{equation}
Apparently, $X=\prod_n \mu_{n,n+1}^z$ remains a symmetry for $H_{\rm LG}(\pi)$. However, the last term changes sign under $X_{\rm even}$. Therefore we need to redefine $X_{\rm even}$ as 
\begin{equation}
    X'_{\rm even}=\mu^x_{N,1}X_{\rm even},
\end{equation}
which then implies $X'_{\rm even}X=-XX'_{\rm even}$.

\end{document}